\documentstyle[aps,prl,epsf,floats,multicol,amssymb,tighten]{revtex}
\begin{document}

\newcommand{\nonb}{\nonumber}
\newcommand{\RA}{\rangle}
\newcommand{\LA}{\langle}
\newcommand{\LL}{\langle \langle}
\newcommand{\RR}{\rangle \rangle}
\newcommand{\HG}{\hat{G}}
\newcommand{\HK}{\hat{K}}
\newcommand{\Ht}{\hat{t}}
\newcommand{\HA}{\hat{A}}
\newcommand{\HI}{\hat{I}}
\newcommand{\HX}{\hat{X}}
\newcommand{\HY}{\hat{Y}}
\newcommand{\HL}{\hat{L}}
\newcommand{\HM}{\hat{M}}
\newcommand{\HN}{\hat{N}}
\newcommand{\Hg}{\hat{g}}
\newcommand{\Hrho}{\hat{\rho}}
\newcommand{\DA}{{\cal D}^A}
\newcommand{\DR}{{\cal D}^R}
\newcommand{\CG}{{\cal G}}
\newcommand{\D}{{\cal D}}
\newcommand{\W}{{{\cal W}_{n,m}(N_a,N_b)}}
\newcommand{\vk}{{\vec{k}}}
\newcommand{\vx}{{\vec{x}}}
\newcommand{\DOSINV}{\frac{m a_0^2}{2 \pi^2 \hbar^2}}
\newcommand{\DOS}{\frac{2 \pi^2 \hbar^2}{ m a_0^2}}

\newenvironment{tab}[1]
{\begin{tabular}{|#1|}\hline}
{\hline\end{tabular}}

\newcommand{\fig}[2]{\epsfxsize=#1\epsfbox{#2}} \reversemarginpar 
\bibliographystyle{prsty}

\title{Microscopic theory of non local pair correlations
in metallic F/S/F trilayers}
\author{V. Apinyan and R. M\'elin\thanks{melin@polycnrs-gre.fr}}
\address{Centre de Recherches sur les Tr\`es Basses
Temp\'eratures (CRTBT)\thanks{U.P.R. 5001 du CNRS,
Laboratoire conventionn\'e avec l'Universit\'e Joseph Fourier}\\
{CNRS BP 166X, 38042 Grenoble Cedex, France}}
\maketitle

\begin{abstract}
We consider a microscopic theory of F/S/F trilayers
with metallic or insulating ferromagnets.
The
trilayer with metallic ferromagnets is controlled by
the formation of non local
pair correlations among the two ferromagnets which
do not exist with insulating ferromagnets.
The difference between the insulating
and ferromagnetic models can
be understood from lowest order diagrams.
Metallic ferromagnets are
controlled by non local pair correlations and
the superconducting gap is larger if the
ferromagnetic electrodes have a
parallel spin orientation. Insulating
ferromagnets are controlled by
pair breaking and the 
superconducting gap is smaller if the
ferromagnetic electrodes have a
parallel spin orientation.
The same behavior is 
found in the presence of
disorder in the microscopic phase variables
and also in the presence of a partial spin
polarization of the ferromagnets.
The different behaviors of the metallic
and insulating trilayers may be probed in
experiments.
\end{abstract}

\widetext


\section{Introduction}

Spin polarized quantum transport has focussed an important
interest recently. One of the challenges in
this field of research is to manipulate correlated pairs of electrons
in solid state
devices.
Several possibilities have been proposed
recently~\cite{Martin,Loss,Feinberg,Falci,Melin,Melin-Feinberg},
some of which involve superconductivity and magnetism.

There is a rich physics occurring at a single
F/S interface. For instance it is
well established that
the superconducting order
parameter induced in a ferromagnetic metal
can oscillate in
space~\cite{Fulde-Ferrel,Larkin,Clogston,Demler}.
In S/F/S Josephson junctions these oscillations
can induce a change of sign in the Josephson
coupling~\cite{Buzdin1}.
This gives rise to the $\pi$-state that
has been probed recently in two
experiments~\cite{Ryazanov,Kontos}.
Another effect taking place at F/S interfaces
is the suppression
of Andreev reflection by spin
polarization~\cite{deJong}. This
has been probed in recent transport
experiments, either with highly
transparent point contacts~\cite{Soulen}
or with intermediate interface
transparencies~\cite{Upadhyay}.
Other systems such as F/S interfaces in
diffusive heterostructures have been
the subject of several experimental
investigations~\cite{Petrashov1,Lawrence,Vasko,Giroud,Petrashov2,Filip}.
These works 
have generated many theoretical discussions
(see for instance
Refs.~\cite{Falko,Jedema,Imamura,Zutic,Kashiwaya,Zhu99,Fazio,Melin-Euro}).

The specific features associated to
transport in multiterminal systems have been
discussed recently
with various methods such as Landauer formalism~\cite{Feinberg},
lowest order perturbation for low transparency
interfaces~\cite{Falci} or
non perturbative solutions for high transparency
interfaces~\cite{Melin-Feinberg}.
It was shown theoretically that
the conductance associated to
Andreev reflection is equal to
the conductance associated to elastic
cotunneling~\cite{Falci,Melin-Feinberg}.
This could be probed
in future experiments by measuring the conductance
as a function of the relative spin orientation of
the ferromagnetic electrodes.

It is also important to understand 
equilibrium properties in multiterminal hybrid systems.
The proximity effect at
F/S interfaces has been discussed in details
recently~\cite{Zhu,Halterman}.
It is well established theoretically
that there exists
oscillations of the critical
temperature in F/S multilayers as the exchange
field and thickness of the F layer are varied~\cite{Buzdin2}.
These oscillations of the critical temperature
have been probed experimentally in several systems:
Nb/Gd multilayers~\cite{exp1,exp2}, Nb/CuMn
multilayers~\cite{exp3}, Nb/Gd/Nb trilayers~\cite{exp4}
and Fe/Nb/Fe trilayers~\cite{exp5}.
Usadel equations have been applied recently to
discuss diffusive F/S/F trilayers~\cite{Buzdin}.
F/S/F trilayers have also been discussed
recently in connection with possible
device applications such as a
superconducting magnetoresistive memory
elements~\cite{Oh} or
a superconducting spin switch~\cite{Tagirov}.
The physics of the
F/S/F trilayer with insulating ferromagnets
is controlled by pair
breaking~\cite{deGennes,Deutscher,Hauser}.
Single electron states in
the superconductor are coupled to an effective
exchange field that cancels
if the two ferromagnets have an antiparallel spin
orientation.
As a consequence the
superconducting gap is
smaller if the ferromagnets have a parallel
spin orientation.

F/S/F trilayers with metallic 
ferromagnets have been investigated in a 
recent work~\cite{Melin-JPC} on the basis
of effective Green's functions.
It was found that the physics is not controlled
by pair breaking, contrary to the F/S/F trilayer
with insulating ferromagnets.
It was found that with metallic ferromagnets
the superconducting
gap is larger if the ferromagnetic electrodes
have a parallel spin orientation~\cite{Melin-JPC}.
It was proposed that
the qualitative physics of multiterminal devices
can be characterized by linear
superpositions of pair states~\cite{Melin-JPC}.
We can therefore contrast two different 
situations:
\begin{itemize}
\item[(i)] F/S/F trilayers with
insulating ferromagnets are controlled by single
electron states. The superconducting
order parameter is smaller if the
ferromagnetic electrodes have a parallel
spin orientation.
\item[(ii)] F/S/F trilayers with metallic ferromagnets
are controlled by non local pair correlations.
The superconducting
order parameter is smaller if the ferromagnetic
electrodes have an antiparallel
spin orientation.
\end{itemize}

The goal of our article is to discuss
F/S/F trilayers with
metallic and insulating ferromagnets, as well
as a ``mixed'' trilayer with an insulating and a
metallic ferromagnet. We use two different
approaches, either analytical (with some
approximations)
or based on exact diagonalizations.

The article is organized as follows.
The model is given in section~\ref{sec:prelim},
as well as technical preliminaries.
Half-metal ferromagnets are
discussed in section~\ref{sec:half-metal}.
This discussion is extended in section~\ref{sec:h-ex}
to describe an arbitrary exchange field.
We present exact diagonalizations 
in section~\ref{sec:num}.
Final remarks are given in section~\ref{sec:conclu}.

\section{Preliminaries}
\label{sec:prelim}
\subsection{The model}
\label{sec:themodel}
We consider throughout the article
a superconductor in contact
with several ferromagnetic electrodes. The superconductor
is three dimensional but a one dimensional
geometry will also be used in the numerical simulations.
We describe the superconductor by a tight binding BCS
model in which the electrons can hop between
neighboring ``sites'' on a square lattice
having a lattice parameter $a_0$. 
The BCS Hamiltonian takes the form
\begin{equation}
\label{eq:tight}
{\cal H}_{\rm BCS} = \sum_{ \langle \alpha,\beta \rangle,\sigma}
-t \left( c_{\alpha,\sigma}^+ c_{\beta,\sigma} +
c_{\beta,\sigma}^+ c_{\alpha,\sigma} \right)
+ \sum_{\alpha} \left( \Delta_{\alpha} 
c_{\alpha,\uparrow}^+ c_{\alpha , \downarrow}^+ +
\Delta_{\alpha}^*
c_{\alpha , \downarrow} c_{\alpha,\uparrow} \right)
,
\end{equation}
where the summation in the kinetic term
is carried out over neighboring
pairs of sites. Without loss of generality, we
assume that the
superconductor conduction band
is half-filled with therefore
$k_F a_0=\pi / 2 $. We note
$\epsilon_F$ the Fermi energy
($\epsilon_F=t$ for a half-filled band)
and we use also the notation $D$ for
the bandwidth.
The physics does not depend
on the details of the band structure. Rather
than using a tight-binding model we can
also use the free electron
dispersion relation $\epsilon(k)
= {\hbar^2 k^2 \over 2 m}$, with
$\epsilon_F= {\hbar^2 k_F^2 \over 2 m}$
the Fermi energy. This dispersion relation
is truncated by a high energy cut-off
$\epsilon(k_{\rm max})=2D=2\epsilon_F$.

The ferromagnetic electrodes are described by the
Stoner model
\begin{equation}
\label{eq:H-Stoner}
{\cal H}_{\rm Stoner} = \sum_{ \langle \alpha,\beta \rangle,\sigma}
-t \left( c_{\alpha,\sigma}^+ c_{\beta,\sigma} +
c_{\beta,\sigma}^+ c_{\alpha,\sigma} \right)
-h_{\rm ex} \sum_{\alpha} \left(
c_{\alpha,\uparrow}^+ c_{\alpha , \uparrow} -
c_{\alpha , \downarrow}^+ c_{\alpha,\downarrow} 
\right)
.
\end{equation}
The case of semi-metal ferromagnets is obtained by
considering that the exchange field $h_{\rm ex}$
is larger than the bandwidth $D$. This model with
no minority-spin conduction channel is discussed in
sections~\ref{sec:half-metal} and~\ref{sec:num}.
The case of partially polarized ferromagnets
corresponding to $h_{\rm ex}<D$ is discussed
in section~\ref{sec:h-ex}.

In the case of half-metal ferromagnets
it will be convenient to use the notation
\begin{equation}
\label{eq:t0-def}
t_0= \frac{t}{\epsilon_F}
\frac{ (a_0 k_F)^2 }{4 \pi}
\end{equation}
in which the hopping matrix element is normalized
with respect to the Fermi energy.
We will use also the notations $\rho^S_0$
for the density of states  in the
superconductor and $\rho^F_\uparrow$ and
$\rho^F_\downarrow$ for the spin-up and spin-down
density of states in the ferromagnetic electrodes:
\begin{eqnarray}
\label{eq:rho-S-0}
\rho^S_0 &=& \frac{1}{\epsilon_F}
\frac{ (a_0 k_F)^2}{4 \pi^2}\\
\label{eq:rho-up}
\rho^F_\uparrow &=& \frac{1}{\epsilon_F^\uparrow}
\frac{ (a_0 k_F^\uparrow)^2}{4 \pi^2}\\
\label{eq:rho-down}
\rho^F_\downarrow &=& \frac{1}{\epsilon_F^\downarrow}
\frac{ (a_0 k_F^\downarrow)^2}{4 \pi^2}
.
\end{eqnarray}
In the case of ferromagnetic metals with two spin
channels (see section~\ref{sec:h-ex}) we will use the
dimensionless parameters
\begin{eqnarray}
\label{eq:x-up}
x_\uparrow &=& \pi^2 t^2 \rho^S_0 \rho_\uparrow\\
x_\downarrow &=& \pi^2 t^2 \rho^S_0 \rho_\downarrow
\label{eq:x-down}
.
\end{eqnarray}

\subsection{The method}
\label{sec:themethod}
We use a Green's function formalism 
(see for
instance~\cite{Melin-Feinberg,Keldysh,Caroli,Cuevas,Martin-Rodero})
to solve the microscopic models given in section~\ref{sec:themodel}.
The first step is to obtain
the expression of the advanced
and retarded Green's functions
$\HG_{i,j}^{A,R}$ in terms of the
advanced and retarded Green's functions of the disconnected
system $\Hg_{i,j}^{A,R}$. This is done by solving the
Dyson equation
\begin{equation}
\label{eq:Dy1}
\HG^{R,A} = \Hg^{R,A} + \Hg^{R,A} \otimes \hat{\Sigma}
\otimes \HG^{R,A}
,
\end{equation}
where the self-energy $\hat{\Sigma}$ contains
all couplings of the tunnel Hamiltonian.
The Green's functions of the connected system
incorporate all excursions
of the electrons in the ferromagnetic electrodes.
The convolution in (\ref{eq:Dy1}) includes
a summation over space labels and a convolution
of times variables. Since we consider a stationary situation,
the latter can be transformed into a product
by Fourier transform.

The advanced Green's
function takes the following form in the Nambu representation:
\begin{equation}
\label{eq:Green-def}
\hat{g}_{\alpha,\beta}^A(t,t') = -i \theta(t-t')
\left( \begin{array}{cc}
\langle \left\{ c_{\alpha,\uparrow}(t) , c_{\beta,\uparrow}^+(t') \right\} \rangle &
\langle \left\{ c_{\alpha,\uparrow}(t) , c_{\beta,\downarrow}(t') \right\} \rangle \\
\langle \left\{ c_{\alpha,\downarrow}^+(t) , c_{\beta,\uparrow}^+(t') \right\} \rangle &
\langle \left\{ c_{\alpha,\downarrow}^+(t) , c_{\beta,\downarrow}(t') \right\} \rangle
\end{array} \right)
,
\end{equation}
where $\alpha$ and $\beta$ are two arbitrary sites in the superconductor.
A similar expression holds for the retarded Green's function.
We adopt the following notation for the Nambu components:
$$
\hat{g}_{\alpha,\beta}^{A,R}(\omega)=
\left( \begin{array}{cc}
g_{\alpha,\beta}^{A,R}(\omega) & f_{\alpha,\beta}^{A,R}(\omega)\\
f_{\alpha,\beta}^{A,R}(\omega) & g_{\alpha,\beta}^{A,R}(\omega)
\end{array} \right)
.
$$
The Nambu representation of the
density of states
$\hat{\rho}_{\alpha,\beta}(\omega) = {1 \over 2 i \pi}
\left[ \hat{g}^A_{\alpha,\beta}(\omega) -
\hat{g}^R_{\alpha,\beta}(\omega) \right]$
will be noted
$$
\hat{\rho}_{\alpha,\beta}(\omega) =
\left( \begin{array}{cc}
\rho_g^{\alpha,\beta}(\omega) & \rho_f^{\alpha,\beta}(\omega) \\
\rho_f^{\alpha,\beta}(\omega) & \rho_g^{\alpha,\beta}(\omega)
\end{array} \right)
,
$$
where
$\rho_g^{\alpha,\beta}(\omega)
= {1 \over 2 i \pi} \left[
g^A_{\alpha,\beta}(\omega)
- g^R_{\alpha,\beta}(\omega) \right]$
and $\rho_f^{\alpha,\beta}(\omega) 
= {1 \over 2 i \pi} \left[
g^A_{\alpha,\beta}(\omega) - 
g^R_{\alpha,\beta}(\omega)
\right]$.
Once the advanced and retarded Green's functions
has been evaluated using~(\ref{eq:Dy1}),
we can evaluate the Keldysh component~\cite{Keldysh}
\begin{equation}
\label{eq:Dy2}
\HG^{+,-} = \left[ \HI + \HG^R \otimes \hat{\Sigma}
\right] \otimes \Hg^{+,-} \otimes
\left[ \HI + \hat{\Sigma} \otimes
\HG^{A} \right]
,
\end{equation}
where $\Hg^{+,-}_{i,j}
= 2 i \pi n_F(\omega-\mu_{i,j})
\hat{\rho}_{i,j}$. 
The Green's function given by~(\ref{eq:Dy2})
can be used either to calculate
transport properties (see for instance~\cite{Caroli})
or to determine the self consistent value of the
superconducting order parameter as we do in
the following (see also~\cite{Martin}).

This method can be used to treat
non local superconducting
pair correlations in the superconductor
and in the ferromagnetic electrodes
(see Ref.~\cite{Melin}).
The pair correlations between two arbitrary
sites $\alpha$ and $\beta$ can be characterized
by the non local Gorkov function 
$\left[G^{+,-}_{\alpha,\beta}(\omega)\right]_{1,2}$.
The local Gorkov function
$\left[G^{+,-}_{\beta,\beta}(\omega)\right]_{1,2}$
can be used to
determine the self consistent value of the superconducting
order parameter
at any site $\beta$ in the superconductor (see~\cite{Tinkham,Martin})
{\sl via} the self-consistency condition
\begin{equation}
\label{eq:self}
\Delta_\beta = -U \int \frac{d \omega}{2 i \pi}
G^{+,-,1,2}_{\beta,\beta}(\omega)
,
\end{equation}
where $U$ is the microscopic attractive interaction.

In the following we concentrate on equilibrium properties.
Namely, the chemical potentials are identical in all
electrodes and there is thus no current flow.
In this situation the Keldysh Green's function~(\ref{eq:Dy2})
simplifies into
\begin{equation}
\label{eq:Gpm-eq}
\HG^{+,-}_{\rm eq} = n_F(\omega-\mu_0) \left( \HG^A - 
\HG^R \right)
,
\end{equation}
where $\mu_0$ is the chemical potential.
The calculations based on~(\ref{eq:Gpm-eq})
will be presented in the main body of the article.
In Appendix~\ref{app:Keldysh} we rederive some of
our results by using directly Eq.~(\ref{eq:Dy2}).

\begin{figure}[thb]
\centerline{\fig{8cm}{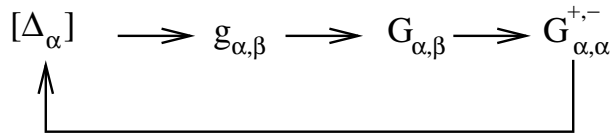}} 
\medskip
\caption{Representation of the successive operations
involved in the calculation of the self consistent
value of the superconducting order parameter.} 
\label{fig:process}
\end{figure}

The self-consistent value of the superconducting
order parameter
can be obtained by iterating
the process on Fig.~\ref{fig:process} which
starts with a uniform gap profile
$[\Delta_\beta]$. From
this gap profile we calculate
the propagator
$g_{\alpha,\beta}(\omega)$
of the superconductor isolated from
the ferromagnetic electrodes.
From Eq.~(\ref{eq:Dy1})
we obtain the
Green's function $G_{\alpha,\beta}^{A,R}$.
From  Eqs.~(\ref{eq:Dy2}) or~(\ref{eq:Gpm-eq})
we deduce the Gorkov
function 
$\left[G^{+,-}_{\alpha,\alpha}(\omega)\right]_{1,2}$,
which is used to recalculate the superconducting order parameter
profile {\sl via} the self consistency
equation~(\ref{eq:self}).

\subsection{The different approaches used
to determine the self consistent gap profile}

\subsubsection{Position of the problem}
\label{sec:position}
To discuss F/S/F trilayers we need
to find reliable determinations
of the self consistent order parameter.
It is in practice impossible
to make an exact analytical calculation
of the self consistent order parameter except
in the limit already considered in Ref.~\cite{Melin-JPC}
where the superconducting gap
is uniform in space (the superconductor is smaller
than the coherence length).
The reason why we cannot find exact solutions
is the following. Let us 
start with a uniform gap profile
$\Delta_\beta \equiv \Delta_0$ and consider the
successive operations on Fig.~\ref{fig:process}.
Because of the contacts between the superconductor
and the ferromagnetic electrodes, the
Green's functions $G^{R,A}_{\alpha,\beta}$
are
not translational invariant. As a result in the next iteration,
the superconducting order parameter is not translational invariant.
The expression of the propagators
of an isolated superconductor
in the presence of a non uniform superconducting order parameter
is not known in general.
The self consistency relation~(\ref{eq:self})
is thus a functional relation:
\begin{equation}
\label{eq:self2}
\label{eq:gap-func}
\Delta_\beta = -U \int \frac{d \omega}{2 i \pi}
G^{+,-,1,2}_{\beta,\beta}([\Delta],\omega)
,
\end{equation}
where the notation $[\Delta]$ means that the right hand
side depends on all values of the gap profile.
As a consequence,
we cannot find exact solutions for the gap profile.

\subsubsection{Local approach}

We present in section~\ref{sec:local}
an approximate analytical 
treatment in which the functional self consistency
relation~(\ref{eq:self2}) is replaced 
by a local relation:
\begin{equation}
\label{eq:self3}
\Delta_\beta = -U \int \frac{d \omega}{2 i \pi}
G^{+,-,1,2}_{\beta,\beta}(\Delta_\beta,\omega)
.
\end{equation}
To transform (\ref{eq:self2}) into (\ref{eq:self3}),
we assume that the
propagators of the isolated superconductor
in the presence of a non uniform
gap has the same energy dependence as
the propagators with a uniform gap. The
propagators $g_{\alpha,\beta}^{A,R}$
and $f_{\alpha,\beta}^{A,R}$
depend on an effective gap
$\Delta_{\alpha,\beta}$.

\subsubsection{Exact diagonalizations}

We present in section~\ref{sec:num}
another possible approach in
which we use exact diagonalizations to solve exactly
the functional form of the
self consistency equation~(\ref{eq:self2}).
This numerical method is restricted to small
system sizes. We will find that the exact diagonalizations
are consistent with the ``local'' approach in
section~\ref{sec:local}
in the sense that we find $\Delta_{\rm AF}
< \Delta_{\rm F}$ with both approaches
for metallic ferromagnets.

\subsection{Green's functions in the presence
of a uniform superconducting order parameter}
\label{sec:unif-Delta}
We end-up this preliminary section by giving the
form of
the Green's function of a superconductor
having a uniform
superconducting order parameter: $\Delta_\beta\equiv\Delta_0$
for all sites $\beta$.
The spectral representation has already been given
in Ref.~\cite{Melin-Feinberg}, as well as the
final form of the propagators below the superconducting
gap. The final form of the propagators above
the superconducting gap is found to be
\begin{eqnarray}
\label{eq:spectral-3D}
\label{eq:Green}
\Hg_{\alpha,\beta}^{R,A}(\omega) &=&
\frac{m a_0^3}{\hbar^2}
\frac{1}{2 \pi |\vx_\alpha-\vx_\beta|}
\exp{\left[ \mp i \psi(\omega) \right]}\\
&&\times
\left\{
\frac{\mp i \sin{\varphi}}{\sqrt{ (\omega-\mu_S)^2 -
\Delta_{\alpha,\beta}^2}}
\left[ \begin{array}{cc}
-(\omega-\mu_S) &  \Delta_{\alpha,\beta} \\
 \Delta_{\alpha,\beta} & -(\omega-\mu_S) \end{array}
\right]
- \cos{\varphi}
\left[ \begin{array}{cc}
1 & 0 \\ 0 & 1
\end{array} \right] \right\}
,\nonb
\end{eqnarray}
where the phase in the prefactor is given by
\begin{equation}
\label{eq:phase}
\psi(\omega) = {1 \over v_F}
|\vx_\alpha-\vx_\beta| \sqrt{(\omega-\mu_S)^2-
\Delta_{\alpha,\beta}^2}
,
\end{equation}
and $\varphi(\omega)=k_F |\vx_\alpha-\vx_\beta|$.
If $\omega \gg \Delta$, the Green's function
reduces to
$$
g_{\alpha,\beta}^A(\omega) = -
{m a_0^3 \over \hbar^2} 
{1 \over 2 \pi R_{\alpha,\beta}}
\exp{ \left[ i \left( \psi(\omega) +
\varphi(\omega) \right) \right]}
,
$$
which will be used in section~\ref{sec:phases}.

\section{Half-metal ferromagnets and ferromagnets
with both spin channels: analytical results}
\label{sec:local}
\label{sec:half-metal}

We discuss in this section mainly the solutions of
F/S/F trilayers with half-metal ferromagnets
having a single spin conduction channel.
In section~\ref{sec:h-ex} 
we extend our discussion to a model having 
both spin conduction channels.
As discussed in section~\ref{sec:position},
the self-consistency equation
for the superconducting order parameter~(\ref{eq:gap-func})
is a functional of the gap profile. In this section,
we replace the functional equation~(\ref{eq:gap-func})
by the local equation~(\ref{eq:self3}).

\subsection{Superconductor connected to a single-channel
half-metal ferromagnet}
\label{sec:(I)}
\begin{figure}[thb]
\centerline{\fig{7cm}{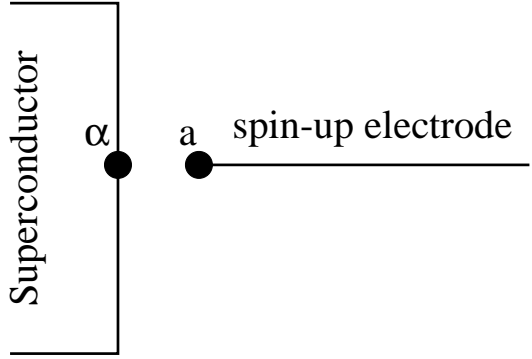}} 
\medskip
\caption{Representation of a model in which a single
channel spin-up electrode is in contact with a
superconductor. 
}
\label{fig:1channel}
\end{figure}
\begin{figure}[thb]
\centerline{\fig{10cm}{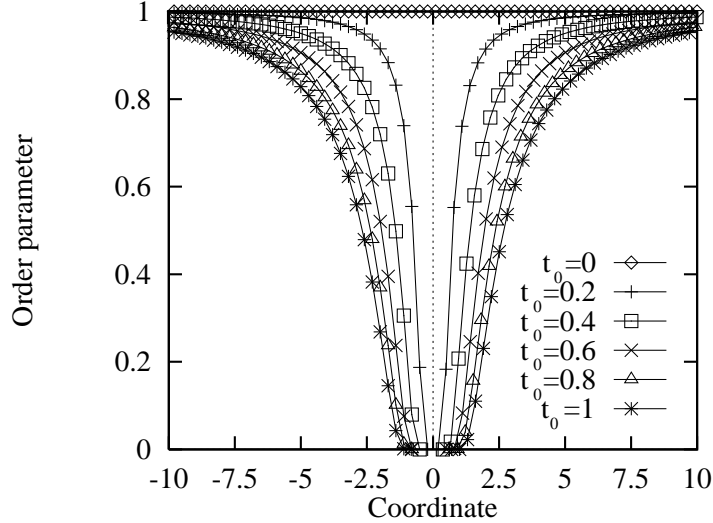}} 
\medskip
\caption{Variation of the superconducting order parameter
with a single ferromagnetic channel.
We used realistic parameters:
the Fermi energy is $\epsilon_F = 10$~eV and the value of
the attractive electron~--~electron interaction
is such that the bulk superconducting order parameter
is $\Delta_{\rm bulk}=1$~meV.} 
\label{fig:gap-1chan}
\end{figure}
We first consider the case on Fig.~\ref{fig:1channel}
where a superconductor is connected to a 
single-channel half-metal ferromagnetic electrode.
Using the Dyson equation~(\ref{eq:Dy1})
and the expression~(\ref{eq:Gpm-eq}) of the equilibrium
Gorkov function, we obtain easily the local Gorkov function
\begin{equation}
\label{eq:Gorkov-local-1ch}
G^{+,-}_{\beta,\beta} = 2 i \pi n_F(\omega)
\left\{ \rho^{\beta,\beta}_f + 
\frac{ |t_{a,\alpha}|^2}{2 i \pi}
\left[ \frac{1}{\cal D^A} g^{a,a,A}_{1,1}
g^{\beta,\alpha,A} f^{\alpha,\beta,A}
- \frac{1}{\cal D^R} g^{a,a,R}_{1,1}
g^{\beta,\alpha,R} f^{\alpha,\beta,R}
\right] \right\}
,
\end{equation}
with
\begin{equation}
\label{eq:D-1contact}
{\cal D}=1-|t^{a,\alpha}|^2 g^{a,a}_{1,1}
g^{\alpha,\alpha}
.
\end{equation}

\subsubsection{Gap profile with fixed phases}

We first make the additional
assumption that the electronic phase in
the Green's function~(\ref{eq:Green}) does not
depend on distance:
$\varphi=-\pi/2$ for all distances. 
Phase averaging will be discussed in section~\ref{sec:phases}.
The gap profile is found to be
\begin{equation}
\label{eq:gap-1channel}
\Delta_\beta = 2 D \exp{\left\{ - \frac{1}{U}
\DOS
\left[ 1 - \left({a_0 \over R_{\alpha,\beta}
}\right)^2
\frac{t_0^2}{1 + t_0^2}  \right]^{-1} \right\}}
,
\end{equation}
where $t_0$ is given by Eq.~(\ref{eq:t0-def}).
$R_{\alpha,\beta}=|\vx_\alpha-\vx_\beta|$
is the distance between sites $\alpha$
and $\beta$ in the superconductor.
Far away from the contact the superconducting order
parameter is equal to the bulk value.
The minimum value of the superconducting order parameter
at the contact
can be estimated from Eq.~(\ref{eq:gap-1channel}) by
replacing $R_{\alpha,\beta}$ by the lattice spacing $a_0$:
$$
\Delta_\alpha = 2 D \exp{\left\{ - \frac{1}{U}
\DOS
\left[ 1 + t_0^2 \right] \right\} }
.
$$
The complete gap profile is shown on Fig.~\ref{fig:gap-1chan}
for several values of the hopping between the superconductor
and the ferromagnetic channel.

\subsubsection{Role of $2 k_F$ oscillations}
\label{sec:2kf-local}
\begin{figure}[thb]
\centerline{\fig{10cm}{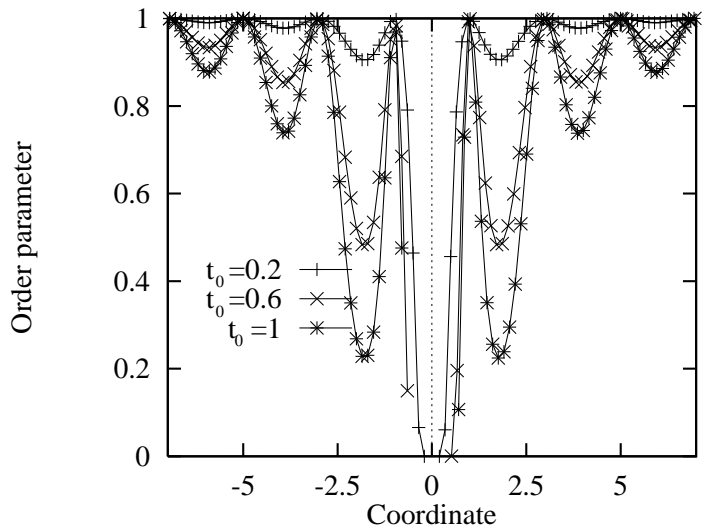}} 
\medskip
\caption{Variation of the superconducting order parameter as a function
of the distance to the contact in the case
of a superconductor connected to a single-channel
ferromagnetic electrode.
We have incorporated the oscillatory phase factor
in~(\ref{eq:gap-1chan-osc}) and we suppose
that $k_F a_0={\pi \over 2}$. The period of the oscillations
is thus $2 a_0$, as expected for $2 k_F$ oscillations.
The parameters are the same
as on Fig.~\ref{fig:gap-1chan}.} 
\label{fig:gap-1chan-osc}
\end{figure}

Now we discuss the role played by the phase
$\varphi$
appearing in the Green's function $g_{\alpha,\beta}$
(see Eq.(\ref{eq:Green})).
In the presence
of this phase factor, the self consistent superconducting order parameter
develops $2 k_F$ oscillations:
\begin{equation}
\label{eq:gap-1chan-osc}
\Delta_\beta = 2 D \exp{\left\{ - \frac{1}{U}
\DOS
\left[ 1 - \left({a_0 \over R_{\alpha,\beta}
}\right)^2
\frac{t_0^2}{1 + t_0^2} 
\sin^2{\left(k_F R_{\alpha,\beta} \right)}
\right]^{-1} \right\}}
,
\end{equation}
where $R_{\alpha,\beta} = |\vx_\alpha - \vx_\beta|$.
The gap profile is shown on Fig.~\ref{fig:gap-1chan-osc}.
One may notice that $\Delta_{\rm bulk}-
\Delta(R_{\alpha,\beta})$ deduced from Fig.~\ref{fig:gap-1chan-osc}
is related to the wave function of a spin-up
electron injected at site $\alpha$ in the superconductor.
Namely the superconducting gap is maximal when the
spin-up wave function is minimal.

\subsection{Superconductor connected to two 
single-channel half-metal ferromagnets}
\label{sec:(III)}
\begin{figure}[thb]
\centerline{\fig{7cm}{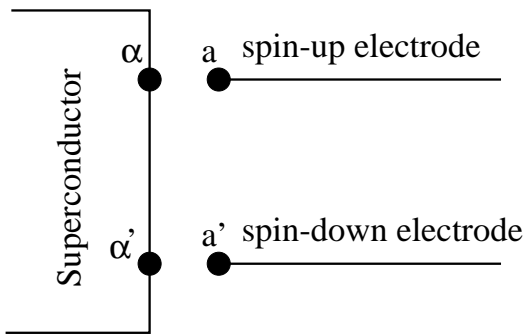}} 
\medskip
\caption{The model in which two single
channel half-metal electrodes are in contact with a
superconductor.}
\label{fig:2channel} 
\end{figure}
Now we consider that two single-channel half-metal
ferromagnets are connected to a superconductor
(see Fig.~\ref{fig:2channel}).
We assume that the electronic phases are fixed
to the value $\varphi=-\pi/2$ for all 
distances and postpone for section~\ref{sec:phases}
the discussion of phase averaging.

\subsubsection{Antiferromagnetic alignment}
\label{sec:main-AF}
Let us consider the model on Fig.~\ref{fig:2channel}
in which two single-channel half-metal
spin-up and spin-down electrodes
are in contact with a superconductor.
The local Green's function takes the form
\begin{equation}
\label{eq:G-loc}
\HG^{\beta,\beta} = \Hg^{\beta,\beta}
+ \Hg^{\beta,\alpha} \Ht^{\alpha,a} \HG^{a,\beta}
+ \Hg^{\beta,\alpha'} \Ht^{\alpha',a'} \HG^{a',\beta}
.
\end{equation}
The propagators $\HG^{a,\beta}$ and $\HG^{a',\beta}$
are given by
\begin{eqnarray}
\label{eq:G-debut-A}
\HG^{a,\beta} &=& t^{a,\alpha} g^{a,a}_{1,1}
\left[\begin{array}{cc}
\tilde{g}^{\alpha,\beta} & \tilde{f}^{\alpha,\beta} \\
0 & 0
\end{array} \right]\\
\HG^{a',\beta} &=& -t^{a',\alpha'} g^{a',a'}_{2,2}
\left[\begin{array}{cc}
0 & 0 \\
\tilde{f}^{\alpha',\beta} & \tilde{g}^{\alpha',\beta} \\
\end{array} \right]
,
\end{eqnarray}
where
\begin{eqnarray}
\label{eq:tildeg1}
\tilde{g}^{\alpha,\beta} &=& \frac{1}{{\cal D}_{\rm AF}}
\left[ g^{\alpha,\beta} + |t^{a',\alpha'}|^2
g^{a',a'}_{2,2} \left( f^{\alpha,\alpha'}
f^{\alpha'\beta} - g^{\alpha',\alpha'}
g^{\alpha,\beta} \right) \right]\\
\tilde{f}^{\alpha,\beta} &=& \frac{1}{{\cal D}_{\rm AF}}
\left[ f^{\alpha,\beta} + |t^{a',\alpha'}|^2
g^{a',a'}_{2,2} \left( f^{\alpha,\alpha'}
g^{\alpha',\beta} - g^{\alpha',\alpha'}
f^{\alpha,\beta} \right) \right]
\label{eq:G-fin-A}
,
\end{eqnarray}
and where
\begin{equation}
\label{eq:DAF-2contact}
{\cal D}_{\rm AF} = \left[ 1 - |t^{a,\alpha}|^2
g^{a,a}_{1,1} g^{\alpha,\alpha} \right]
\left[ 1 - |t^{a',\alpha'}|^2
g^{a',a'}_{2,2} g^{\alpha',\alpha'} \right] -
|t^{a,\alpha}|^2 |t^{a',\alpha'}|^2
g^{a,a}_{1,1} g^{a',a'}_{2,2}
f^{\alpha,\alpha'} f^{\alpha',\alpha}
.
\end{equation}
The self-consistent superconducting order parameter
is obtained by evaluating the high energy behavior
of the local Gorkov function given by~(\ref{eq:Gpm-eq}).
We use a ``local'' approximation in which the
gaps $\Delta_{\alpha,\beta}$ and
$\Delta_{\alpha',\beta}$ appearing in the
propagators $g_{\alpha,\beta}$ and $f_{\alpha,\beta}$
are replaced by the local gap $\Delta_\beta$.
The high energy behavior of the local Gorkov
is found to be
\begin{equation}
\label{eq:Gorkov-AF}
\HG^{+,-}_{\beta,\beta} = - 2 i \pi n_F(\omega)
\DOSINV 
\left(\frac{\Delta_\beta}{\omega}\right)
\Lambda^{\rm AF}_{\rm Metal}
,
\end{equation}
with
\begin{equation}
\label{eq:Lambda-AF}
\Lambda^{\rm AF}_{\rm Metal} = 1 - {a_0^2 \over R_{\alpha,\beta}^2}
\left(
\frac{ |t^{a,\alpha}_0|^2}{ 1 + |t^{a,\alpha}_0|^2}
\right)
- {a_0^2  \over R_{\alpha',\beta}^2}
\left(
\frac{ |t^{a',\alpha'}_0|^2}{ 1 + |t^{a',\alpha'}_0|^2}
\right)
+ \frac{a_0^3}{R_{\alpha,\beta} R_{\alpha',\beta}
R_{\alpha,\alpha'}}
\left(
\frac{ |t^{a,\alpha}_0|^2}{ 1 + |t^{a,\alpha}_0|^2}
\right)
\left(
\frac{ |t^{a',\alpha'}_0|^2}{ 1 + |t^{a',\alpha'}_0|^2}
\right)
,
\end{equation}
where $t^{a,\alpha}_0$
and $t^{a',\alpha'}_0$ are the tunnel matrix
elements normalized to the Fermi energy
(see Eq.~(\ref{eq:t0-def})).

\subsubsection{Ferromagnetic alignment}
Using the same method for the ferromagnetic
alignment, we obtain the high energy behavior of
the local Gorkov function
\begin{equation}
\label{eq:Gorkov-F}
\HG^{+,-}_{\beta,\beta} =- 2 i \pi n_F(\omega)
\DOSINV 
\left(\frac{\Delta_\beta}{\omega}\right)
\Lambda^{\rm F}_{\rm Metal}
,
\end{equation}
with
\begin{equation}
\label{eq:A}
\Lambda^{\rm F}_{\rm Metal} = 1 - {a_0^2 \over R_{\alpha,\beta}^2}
\left(
\frac{ |t^{a,\alpha}_0|^2}{ 1 + |t^{a,\alpha}_0|^2}
\right)
- {a_0^2 \over R_{\alpha',\beta}^2}
\left(
\frac{ |t^{a',\alpha'}_0|^2}{ 1 + |t^{a',\alpha'}_0|^2}
\right)
+ 2 |t^{a,\alpha}_0|^2 |t^{a',\alpha'}_0|^2
{1 \over {\cal D}_F}
\frac{a_0^3}{R_{\alpha,\beta} R_{\alpha',\beta}
R_{\alpha,\alpha'}}
,
\end{equation}
and
\begin{equation}
\label{eq:D-F}
{\cal D}_{\rm F} = \left[ 1 - |t^{a,\alpha}|^2
g^{a,a}_{1,1} g^{\alpha,\alpha} \right]
\left[ 1 - |t^{a',\alpha'}|^2
g^{a',a'}_{2,2} g^{\alpha',\alpha'} \right] -
|t^{a,\alpha}|^2 |t^{a',\alpha'}|^2
g^{a,a}_{1,1} g^{a',a'}_{2,2}
g^{\alpha,\alpha'} g^{\alpha',\alpha}
.
\end{equation}
To order $1/R^3$, Eq.~(\ref{eq:A}) becomes
\begin{equation}
\label{eq:Lambda-F-fin}
\Lambda^{\rm F}_{\rm Metal} = 1
- {a_0^2 \over R_{\alpha,\beta}^2}
\left(
\frac{ |t^{a,\alpha}_0|^2}{ 1 + |t^{a,\alpha}_0|^2}
\right)
- {a_0^2  \over R_{\alpha',\beta}^2}
\left(
\frac{ |t^{a',\alpha'}_0|^2}{ 1 + |t^{a',\alpha'}_0|^2}
\right)
+ 2 \frac{a_0^3}{R_{\alpha,\beta} R_{\alpha',\beta}
R_{\alpha,\alpha'}}
\left(
\frac{ |t^{a,\alpha}_0|^2}{ 1 + |t^{a,\alpha}_0|^2}
\right)
\left(
\frac{ |t^{a',\alpha'}_0|^2}{ 1 + |t^{a',\alpha'}_0|^2}
\right)
.
\end{equation}
Comparing Eqs.~(\ref{eq:Lambda-AF}) and~(\ref{eq:Lambda-F-fin}),
we see that:
\begin{itemize}
\item[(i)] As expected, the ``local'' contributions
of order $1/R_{\alpha,\beta}^2$ and
$1/R_{\alpha',\beta}^2$ do not depend on the relative
spin orientation of the ferromagnetic electrodes.
The local contributions generate a reduction
of the superconducting order parameter.
\item[(ii)] The lowest order ``non local'' contribution
arises at order $1/ [ R_{\alpha,\beta} R_{\alpha',\beta}
R_{\alpha,\alpha'}]$, and depends on the relative
spin orientation of the ferromagnetic electrodes
{\sl via} a factor of two in Eq.~(\ref{eq:Lambda-F-fin}),
not present in Eq.~(\ref{eq:Lambda-AF}).
Because of this non local contribution, the superconducting
gap is larger in the ferromagnetic alignment,
which can receive a simple interpretation
in terms of the diagrams contributing to this non
local term
(see section~\ref{sec:Feynman}).
\end{itemize}

\subsubsection{Gap profiles}
\begin{figure}[thb]
\centerline{\fig{10cm}{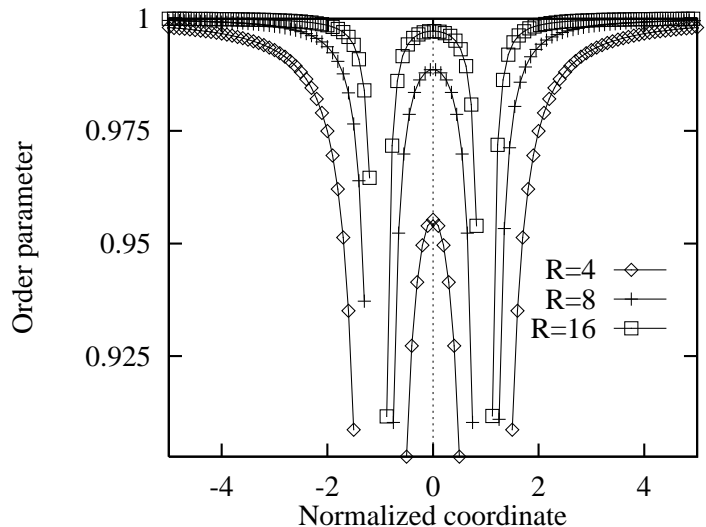}} 
\medskip
\caption{Variation of the superconducting order parameter
in the presence of two ferromagnetic electrodes.
It is assumed that the point
$\beta$ is aligned with the points $\alpha$ and $\alpha'$.
The coordinate is normalized
to the separation $R_{\alpha,\alpha'}$ between the contacts. 
The different curves correspond to
$R_{\alpha,\alpha'}=4$~($\Diamond$),
$R_{\alpha,\alpha'}=8$~($+$) and
$R_{\alpha,\alpha'}=16$~($\Box$).
We used the same
parameters as on Fig.~\ref{fig:gap-1chan}.
The contacts have a low
transparency: $t^{a,\alpha}_0=t^{a',\alpha'}_0=0.1$.
The ferromagnetic electrodes have an antiparallel
spin orientation.} 
\label{fig:gap-2chan-tun}
\end{figure}
\begin{figure}[thb]
\centerline{\fig{10cm}{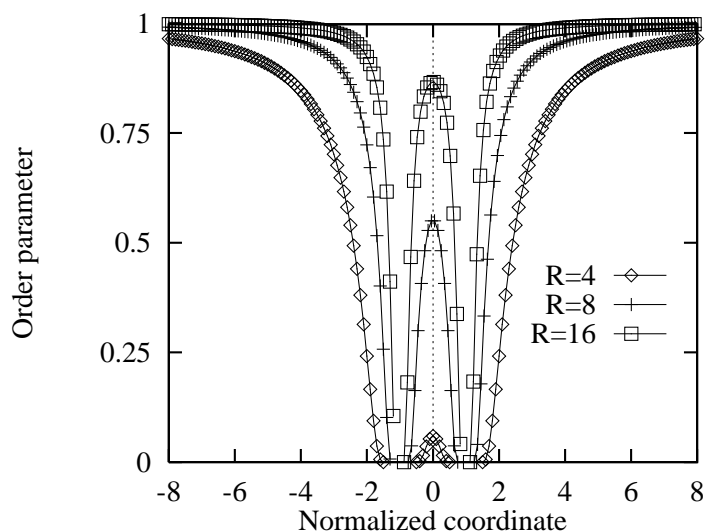}} 
\medskip
\caption{The same as Fig.~\ref{fig:gap-2chan-tr}
with high transparency contacts:
$t^{a,\alpha}_0=t^{a',\alpha'}_0= 1$.} 
\label{fig:gap-2chan-tr}
\end{figure}

The gap profiles are shown on Fig.~\ref{fig:gap-2chan-tun}
in the tunnel regime and Fig.~\ref{fig:gap-2chan-tr}
in the high transparency regime. The gap is reduced
close to the contacts with the ferromagnets,
which was already obtained for the single
channel model in sections~\ref{sec:(I)}
and~\ref{sec:2kf-local} (see Figs.~\ref{fig:gap-1chan}
and~\ref{fig:gap-1chan-osc}).

\subsubsection{Difference between the superconducting gaps
in the ferromagnetic and antiferromagnetic alignments}
\begin{figure}[thb]
\centerline{\fig{10cm}{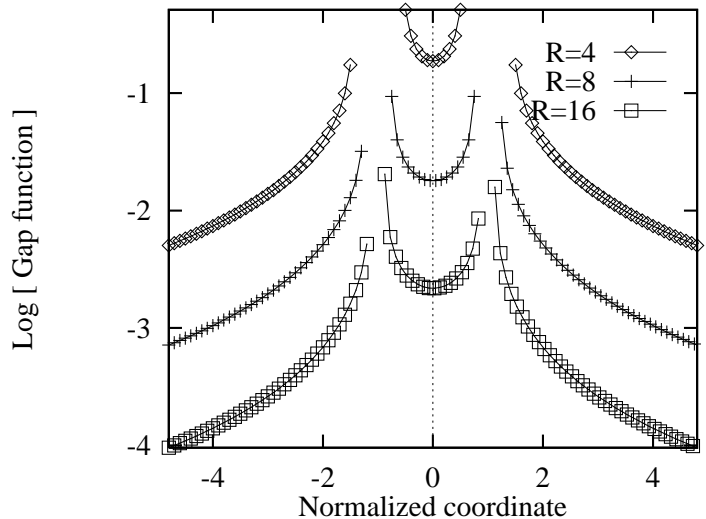}} 
\medskip
\caption{ Variation of the logarithm of 
$\delta_\beta$ defined by~(\ref{eq:gap-ratio-def}).
The ferromagnetic gap is larger than the
antiferromagnetic gap.
The parameters are the same as on Fig.~\ref{fig:gap-1chan}.
The contacts have a high
transparency: $t^{a,\alpha}_0=t^{a',\alpha'}_0= 1$. } 
\label{fig:gap-ratio}
\end{figure}
At each point $\beta$ in the superconductor, we
define $\delta_\beta$ as
\begin{equation}
\label{eq:gap-ratio-def}
\delta_\beta = 2 \frac{\Delta^{\rm F}_\beta -
\Delta^{\rm AF}_\beta}
{ \Delta^{\rm F}_\beta +
\Delta^{\rm AF}_\beta}
.
\end{equation}
$\delta_\beta$ is positive for metallic ferromagnets,
and takes a simple form at large distance:
\begin{equation}
\label{eq:scaling1}
\delta_\beta \simeq \frac{1}{U} \DOS
\frac{ |t^{a,\alpha}_0|^2}{ 1 + |t^{a,\alpha}_0|^2}
\frac{ |t^{a',\alpha'}_0|^2}{ 1 + |t^{a',\alpha'}_0|^2}
\frac{a_0^3}{R_{\alpha,\beta} R_{\alpha',\beta}
R_{\alpha,\alpha'}}
.
\end{equation}

\subsection{Superconductor connected to two
single-channel insulating ferromagnets}
\label{sec:insulating}

Now we show that we recover the correct physics
in the case of insulating ferromagnets~\cite{deGennes}.
The propagator relevant to describe a ferromagnetic
insulator
decays exponentially with distance and is such
that $g_{i,j}^A=g_{i,j}^R$. The local
propagators $g_{a,a}$ and $g_{a',a'}$ are
real numbers.
Using the
same method as in section~\ref{sec:(III)}
we find that the
Gorkov functions are still given
by~(\ref{eq:Gorkov-AF}) and~(\ref{eq:Gorkov-F})
but with a different form of $\Lambda^{\rm AF}$
and $\Lambda^{\rm F}$. The expression of the
Gorkov functions to order $1/R^3$ is the following:
\begin{eqnarray}
\label{eq:Lambda-AF-In}
\Lambda^{\rm AF}_{\rm Ins} &=& 1 - {a_0^2 \over R_{\alpha,\beta}^2}
\left(
\frac{ |t^{a,\alpha}_0|^4}{ 1 + |t^{a,\alpha}_0|^4}
\right)
- {a_0^2 \over R_{\alpha',\beta}^2}
\left(
\frac{ |t^{a',\alpha'}_0|^4}{ 1 + |t^{a',\alpha'}_0|^4}
\right)\\\nonb
&&- \frac{a_0^3}{R_{\alpha,\beta} R_{\alpha',\beta}
R_{\alpha,\alpha'}}
\left(1 - |t^{a,\alpha}_0|^2 |t^{a',\alpha'}_0|^2
\right)
\left(\frac{|t^{a,\alpha}_0|^2} { 1 + |t^{a,\alpha}_0|^4}\right)
\left(\frac{|t^{a',\alpha'}_0|^2} { 1 + |t^{a',\alpha'}_0|^4}\right)\\
\label{eq:Lambda-F-In}
\Lambda^{\rm F}_{\rm Ins} &=& 1 - {a_0^2  \over R_{\alpha,\beta}^2}
\left(
\frac{ |t^{a,\alpha}_0|^4}{ 1 + |t^{a,\alpha}_0|^4}
\right)
- {a_0^2  \over R_{\alpha',\beta}^2}
\left(
\frac{ |t^{a',\alpha'}_0|^4}{ 1 + |t^{a',\alpha'}_0|^4}
\right)\\\nonb
&&- 2 \frac{a_0^3 }{R_{\alpha,\beta} R_{\alpha',\beta}
R_{\alpha,\alpha'}}
\left(1 - |t^{a,\alpha}_0|^2 |t^{a',\alpha'}_0|^2
\right)
\left(\frac{|t^{a,\alpha}_0|^2} { 1 + |t^{a,\alpha}_0|^4}\right)
\left(\frac{|t^{a',\alpha'}_0|^2} { 1 + |t^{a',\alpha'}_0|^4}\right)
.
\end{eqnarray}
We deduce from~(\ref{eq:Lambda-AF-In})
and~(\ref{eq:Lambda-F-In}) the value of
$\delta_\beta$ defined by~(\ref{eq:gap-ratio-def}):
$$
\delta_\beta=-{1 \over U} \DOS
\left(1 - |t^{a,\alpha}_0|^2 |t^{a',\alpha'}_0|^2
\right)
\frac{|t^{a,\alpha}_0|^2} { 1 + |t^{a,\alpha}_0|^4}
\frac{|t^{a',\alpha'}_0|^2} { 1 + |t^{a',\alpha'}_0|^4}
\frac{a_0^3}{R_{\alpha,\beta} R_{\alpha',\beta}
R_{\alpha,\alpha'}}
.
$$
The values of the normalized tunnel matrix elements
are such that $|t^{a,\alpha}_0| |t^{a',\alpha'}_0|<1$.
As a consequence 
we recover the
perturbative result obtained in Ref.~\cite{deGennes} for
insulating ferromagnets~($\delta_\beta<0$).

\subsection{Phase averaging}
\label{sec:phases}
The microscopic Green's function $\Hg^{\alpha,\beta}$
depends on the phase variables $\varphi_{\alpha,\beta}(\omega)$
and $\psi_{\alpha,\beta}(\omega)$~(see Eq.~(\ref{eq:Green})).
In the preceding subsections,
we have assumed that these phases were fixed to
$\varphi_{\alpha,\beta}(\omega)=-\pi/2$ and
$\psi_{\alpha,\beta}(\omega)=0$ for all distances.
In fact the microscopic phases
phases given by~(\ref{eq:phase})
oscillate rapidly on microscopic scales,
as opposed to the slowly varying prefactor
involving $1/R_{\alpha,\beta}$ in
$\Hg_{\alpha,\beta}$ (see Eq.~(\ref{eq:Green})).
Moreover in a multichannel
model these phases are averaged out
when the summation over all channels is carried out
(see Ref.~\cite{Melin-Feinberg}).
It is thus legitimate
to consider the phases as random
variables
and to average the Gorkov functions over ``phase disorder''. 

\subsubsection{Single-channel problem}
\label{sec:phases-single}
Let us start with the single-channel problem
on Fig.~\ref{fig:1channel}.
The local Gorkov function were
already given in Eq.~(\ref{eq:Gorkov-local-1ch}).
We need to evaluate $\LL g^{\beta,\alpha,A}
f^{\alpha,\beta,A} \RR$ and
$\LL g^{\beta,\alpha,R} f^{\alpha,\beta,R} \RR$,
where $\LL { \mbox{ }} \RR$ denotes the averaging
over phase disorder.
Assuming that the phases are symmetric
(namely $\varphi_{\alpha,\beta}(\omega)
=\varphi_{\beta,\alpha}(\omega)$ and
$\psi_{\alpha,\beta}(\omega)
= \psi_{\beta,\alpha}(\omega)$)
leads to
$\LL g^{\beta,\alpha,A}
f^{\alpha,\beta,A} \RR=0$ and
$\LL g^{\beta,\alpha,R} f^{\alpha,\beta,R} \RR=0$,
from what
we deduce that the superconducting gap does
not depend on the transparency of the contact with
the ferromagnet.
Since this conclusion is not acceptable physically,
we suppose instead that the
phases are antisymmetric:
$\varphi_{\alpha,\beta} = - \pi/2 
+ k_F R_{\alpha,\beta}$,
$\varphi_{\beta,\alpha} = - \pi/2 
- k_F R_{\alpha,\beta}$,
and $\psi_{\alpha,\beta} = - \psi_{\beta,\alpha}$.
Then the expectation value of
$\LL g^{\beta,\alpha,A} f^{\alpha,\beta,A} \RR$
is finite, as it should:
\begin{equation}
\LL g^{\beta,\alpha,A} f^{\alpha,\beta,A} \RR
= {1 \over 2}
\left( {m a_0^2 \over 2 \pi^2 \hbar^2} \right)
\left({a_0 \over R_{\alpha,\beta}} \right)^2
{\Delta_0 \over \sqrt{ \omega^2 - \Delta_0^2}}
,
\end{equation}
from what we deduce the self consistent superconducting
order parameter
\begin{equation}
\label{eq:gap-1chan-av}
\Delta_\beta = 2 D \exp{\left\{ - \frac{1}{U}
\DOS
\left[ 1 - {1 \over 2} \left({a_0 \over R_{\alpha,\beta}
}\right)^2
\frac{t_0^2}{1 + t_0^2} 
\right]^{-1} \right\}}
.
\end{equation}
This form of the gap profile is similar
to Eq.~(\ref{eq:gap-1channel}), except for the
coefficient $1/2$ due to phase averaging.
Now we discuss phase averaging in different types
of two-channel heterostructures.

\subsubsection{Superconductor connected to two
single-channel half-metal ferromagnets}
With two half-metal ferromagnets, we obtain
\begin{eqnarray}
&&\Lambda^{\rm AF}_{\rm Metal} = 1 - {a_0^2 \over R_{\alpha,\beta}^2}
\left(
\frac{ |t^{a,\alpha}_0|^2}{ 1 + |t^{a,\alpha}_0|^2}
\right) \sin^2{(\varphi_{\alpha,\beta})}
- {a_0^2  \over R_{\alpha',\beta}^2}
\left(
\frac{ |t^{a',\alpha'}_0|^2}{ 1 + |t^{a',\alpha'}_0|^2}
\right) \sin^2{(\varphi_{\alpha',\beta})}\\\nonb
&+& \frac{a_0^3}{R_{\alpha,\beta} R_{\alpha',\beta}
R_{\alpha,\alpha'}}
\left(
\frac{ |t^{a,\alpha}_0|^2}{ 1 + |t^{a,\alpha}_0|^2}
\right)
\left(
\frac{ |t^{a',\alpha'}_0|^2}{ 1 + |t^{a',\alpha'}_0|^2}
\right)
\sin{(\varphi_{\alpha,\alpha'})} \cos{\left(\varphi_{\beta,\alpha}
+ \varphi_{\alpha',\beta}\right)}\\
&&\Lambda^{\rm F}_{\rm Metal} = 1 - {a_0^2 \over R_{\alpha,\beta}^2}
\left(
\frac{ |t^{a,\alpha}_0|^2}{ 1 + |t^{a,\alpha}_0|^2}
\right) \sin^2{(\varphi_{\alpha,\beta})}
- {a_0^2  \over R_{\alpha',\beta}^2}
\left(
\frac{ |t^{a',\alpha'}_0|^2}{ 1 + |t^{a',\alpha'}_0|^2}
\right) \sin^2{(\varphi_{\alpha',\beta})}\\\nonb
&+& \frac{a_0^3}{R_{\alpha,\beta} R_{\alpha',\beta}
R_{\alpha,\alpha'}}
\left(
\frac{ |t^{a,\alpha}_0|^2}{ 1 + |t^{a,\alpha}_0|^2}
\right)
\left(
\frac{ |t^{a',\alpha'}_0|^2}{ 1 + |t^{a',\alpha'}_0|^2}
\right) \left\{
\cos{\left(\varphi_{\beta,\alpha} + \varphi_{\alpha,\alpha'}\right)}
\sin{(\varphi_{\alpha',\beta})} +
\cos{\left(\varphi_{\beta,\alpha'} + \varphi_{\alpha',\alpha}\right)}
\sin{(\varphi_{\alpha,\beta})} \right\}
.
\end{eqnarray}
After averaging over phase disorder, we find
\begin{eqnarray}
\label{eq:HALF1}
\LL \Lambda^{\rm AF}_{\rm Metal} \RR &=& 1 - {1 \over 2}
{a_0^2 \over R_{\alpha,\beta}^2}
\left(
\frac{ |t^{a,\alpha}_0|^2}{ 1 + |t^{a,\alpha}_0|^2}
\right) 
- {1 \over 2} {a_0^2  \over R_{\alpha',\beta}^2}
\left(
\frac{ |t^{a',\alpha'}_0|^2}{ 1 + |t^{a',\alpha'}_0|^2}
\right) \\\nonb
&+& {1 \over 2} 
\frac{a_0^3}{R_{\alpha,\beta} R_{\alpha',\beta}
R_{\alpha,\alpha'}}
\left(
\frac{ |t^{a,\alpha}_0|^2}{ 1 + |t^{a,\alpha}_0|^2}
\right)
\left(
\frac{ |t^{a',\alpha'}_0|^2}{ 1 + |t^{a',\alpha'}_0|^2}
\right)\\
\label{eq:HALF2}
\LL \Lambda^{\rm F}_{\rm Metal} \RR &=& 1 - {1 \over 2}
{a_0^2 \over R_{\alpha,\beta}^2}
\left(
\frac{ |t^{a,\alpha}_0|^2}{ 1 + |t^{a,\alpha}_0|^2}
\right) 
- {1 \over 2}
{a_0^2  \over R_{\alpha',\beta}^2}
\left(
\frac{ |t^{a',\alpha'}_0|^2}{ 1 + |t^{a',\alpha'}_0|^2}
\right) \\\nonb
&+& 
\frac{a_0^3}{R_{\alpha,\beta} R_{\alpha',\beta}
R_{\alpha,\alpha'}}
\left(
\frac{ |t^{a,\alpha}_0|^2}{ 1 + |t^{a,\alpha}_0|^2}
\right)
\left(
\frac{ |t^{a',\alpha'}_0|^2}{ 1 + |t^{a',\alpha'}_0|^2}
\right) 
.
\end{eqnarray}
The form of the Gorkov functions is therefore similar to
Eqs.~(\ref{eq:Lambda-AF}) and~(\ref{eq:Lambda-F-fin}), except for the
$1/2$ prefactors. 

\subsubsection{Superconductor connected to two single-channel
insulating ferromagnets}
For a superconductor connected to two
insulating ferromagnets,
we obtain
\begin{eqnarray}
\LL \Lambda^{\rm AF}_{\rm Ins} \RR &=& 1 - {1 \over 2}
{a_0^2 \over R_{\alpha,\beta}^2}
\left(
\frac{ |t^{a,\alpha}_0|^4}{ 1 + |t^{a,\alpha}_0|^4}
\right)
- {1 \over 2} {a_0^2 \over R_{\alpha',\beta}^2}
\left(
\frac{ |t^{a',\alpha'}_0|^4}{ 1 + |t^{a',\alpha'}_0|^4}
\right)\\\nonb
&&- {1 \over 2} \frac{a_0^3}{R_{\alpha,\beta} R_{\alpha',\beta}
R_{\alpha,\alpha'}}
\left(1 - |t^{a,\alpha}_0|^2 |t^{a',\alpha'}_0|^2
\right)
\left(\frac{|t^{a,\alpha}_0|^2} { 1 + |t^{a,\alpha}_0|^4}\right)
\left(\frac{|t^{a',\alpha'}_0|^2} { 1 + |t^{a',\alpha'}_0|^4}\right)\\
\LL \Lambda^{\rm F}_{\rm Ins} \RR
&=& 1 - {1 \over 2} {a_0^2  \over R_{\alpha,\beta}^2}
\left(
\frac{ |t^{a,\alpha}_0|^4}{ 1 + |t^{a,\alpha}_0|^4}
\right)
- {1 \over 2} {a_0^2  \over R_{\alpha',\beta}^2}
\left(
\frac{ |t^{a',\alpha'}_0|^4}{ 1 + |t^{a',\alpha'}_0|^4}
\right)\\\nonb
&&-  \frac{a_0^3 }{R_{\alpha,\beta} R_{\alpha',\beta}
R_{\alpha,\alpha'}}
\left(1 - |t^{a,\alpha}_0|^2 |t^{a',\alpha'}_0|^2
\right)
\left(\frac{|t^{a,\alpha}_0|^2} { 1 + |t^{a,\alpha}_0|^4}\right)
\left(\frac{|t^{a',\alpha'}_0|^2} { 1 + |t^{a',\alpha'}_0|^4}\right)
,
\end{eqnarray}
which differs from~(\ref{eq:Lambda-AF-In})
and~(\ref{eq:Lambda-F-In}) by the $1/2$ coefficients.

\subsubsection{``Mixed'' junction with an insulating and a
metallic single-channel ferromagnet}
Now we consider the ``mixed''
heterostructure on Fig.~\ref{fig:2channel}
in which electrode ``a'' is a single-channel half
metal ferromagnet and electrode ``b''
is insulating. Using the same formalism as in
the preceding sections, we obtain
\begin{eqnarray}
\Lambda^{\rm AF}_{\rm Mixed} &=& 1 - {a_0^2 \over R_{\alpha,\beta}^2}
\left(
\frac{ |t^{a,\alpha}_0|^2}{ 1 + |t^{a,\alpha}_0|^2}
\right) \sin^2{(\varphi_{\alpha,\beta})}\\\nonb
&+& {a_0^2  \over R_{\alpha',\beta}^2}
\left(
\frac{ |t^{a',\alpha'}_0|^2}{ 1 + |t^{a',\alpha'}_0|^4}
\right) \sin{(\varphi_{\beta,\alpha'})}
\left[ \cos{(\varphi_{\alpha'\beta})}
- |t_0^{a',\alpha'}|^2 \sin{(\varphi_{\alpha',\beta}})
\right]
\\\nonb
&+& \frac{a_0^3}{R_{\alpha,\beta} R_{\alpha',\beta}
R_{\alpha,\alpha'}}
\left(
\frac{ |t^{a,\alpha}_0|^2}{ 1 + |t^{a,\alpha}_0|^2}
\right)
\left(
\frac{ |t^{a',\alpha'}_0|^2}{ 1 + |t^{a',\alpha'}_0|^4}
\right)
\sin{(\varphi_{\alpha,\alpha'})}
\left[ \sin{(\varphi_{\beta,\alpha} + \varphi_{\alpha',\beta})}
+ |t_0^{a',\alpha'}|^2 \cos{(\varphi_{\beta,\alpha}
+\varphi_{\alpha',\beta})} \right]\\
\Lambda^{\rm F}_{\rm Mixed} &=& 1 - {a_0^2 \over R_{\alpha,\beta}^2}
\left(
\frac{ |t^{a,\alpha}_0|^2}{ 1 + |t^{a,\alpha}_0|^2}
\right) \sin^2{(\varphi_{\alpha,\beta})}\\\nonb
&+& {a_0^2  \over R_{\alpha',\beta}^2}
\left(
\frac{ |t^{a',\alpha'}_0|^2}{ 1 + |t^{a',\alpha'}_0|^4}
\right) 
\left[ \cos{(\varphi_{\beta,\alpha'})} \sin{(\varphi_{\alpha',\beta})}
-|t_0^{a',\alpha'}|^2 \sin{(\varphi_{\alpha',\beta})}
\sin{(\varphi_{\beta,\alpha'})} \right]\\\nonb
&+& \frac{a_0^3}{R_{\alpha,\beta} R_{\alpha',\beta}
R_{\alpha,\alpha'}}
\left(
\frac{ |t^{a,\alpha}_0|^2}{ 1 + |t^{a,\alpha}_0|^2}
\right)
\left(
\frac{ |t^{a',\alpha'}_0|^2}{ 1 + |t^{a',\alpha'}_0|^4}
\right) \times\\\nonb
&& \left[
\sin{(\varphi_{\beta,\alpha}+\varphi_{\alpha,\alpha'})}
\sin{(\varphi_{\alpha',\beta})}
+ \sin{(\varphi_{\beta,\alpha'}+\varphi_{\alpha',\alpha})}
\sin{(\varphi_{\alpha,\beta})} \right.\\\nonb
&& \left.+ |t_0^{a',\alpha'}|^2 \left(
\cos{(\varphi_{\beta,\alpha}  + \varphi_{\alpha,\alpha'})}
\sin{(\varphi_{\alpha',\beta})}
+ \cos{(\varphi_{\beta,\alpha'} + \varphi_{\alpha',\alpha})}
\sin{(\varphi_{\alpha,\beta})} \right)
\right]
.
\end{eqnarray}
Averaging over phase disorder leads to
\begin{eqnarray}
\label{eq:mixed1}
\LL \Lambda^{\rm AF}_{\rm Mixed} \RR &=& 1 - {1 \over 2}
{a_0^2 \over R_{\alpha,\beta}^2}
\left(
\frac{ |t^{a,\alpha}_0|^2}{ 1 + |t^{a,\alpha}_0|^2}
\right) 
- {1 \over 2} {a_0^2  \over R_{\alpha',\beta}^2}
\left(
\frac{ |t^{a',\alpha'}_0|^4}{ 1 + |t^{a',\alpha'}_0|^4}
\right) \\\nonb
&+& {1 \over 2} 
\frac{a_0^3}{R_{\alpha,\beta} R_{\alpha',\beta}
R_{\alpha,\alpha'}}
\left(
\frac{ |t^{a,\alpha}_0|^2}{ 1 + |t^{a,\alpha}_0|^2}
\right)
\left(
\frac{ |t^{a',\alpha'}_0|^4}{ 1 + |t^{a',\alpha'}_0|^4}
\right)\\
\label{eq:mixed2}
\LL \Lambda^{\rm F}_{\rm Mixed} \RR &=& 1 - {1 \over 2}
{a_0^2 \over R_{\alpha,\beta}^2}
\left(
\frac{ |t^{a,\alpha}_0|^2}{ 1 + |t^{a,\alpha}_0|^2}
\right) \sin^2{(\varphi_{\alpha,\beta})}
+ {1 \over 2}
{a_0^2  \over R_{\alpha',\beta}^2}
\left(
\frac{ |t^{a',\alpha'}_0|^4}{ 1 + |t^{a',\alpha'}_0|^4}
\right)\\\nonb
&+& \frac{a_0^3}{R_{\alpha,\beta} R_{\alpha',\beta}
R_{\alpha,\alpha'}}
\left(
\frac{ |t^{a,\alpha}_0|^2}{ 1 + |t^{a,\alpha}_0|^2}
\right)
\left(
\frac{ |t^{a',\alpha'}_0|^4}{ 1 + |t^{a',\alpha'}_0|^4}
\right) 
.
\end{eqnarray}
As a consequence this heterostructure
behaves like the full metallic heterostructure
($\delta_\beta>0$).

\subsection{Lowest order diagrams}
\label{sec:Feynman}
In this section we point out
a simple rule that can be used to determine
whether $\Delta_\beta^{\rm F}>\Delta_\beta^{\rm AF}$
or whether
$\Delta_\beta^{\rm F}<\Delta_\beta^{\rm AF}$.
\begin{figure}[thb]
\centerline{\fig{6cm}{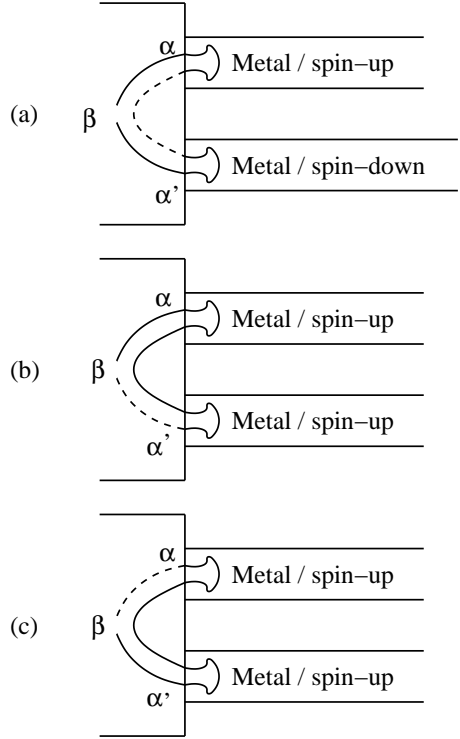}} 
\medskip
\caption{Lowest order processes in
the case of metallic ferromagnets. (a) corresponds
to~(\ref{eq:proc1}), (b) corresponds to~(\ref{eq:proc2}),
and (c) corresponds to~(\ref{eq:proc3}).} 
\label{fig:diag1}
\end{figure}
Let us start with the metallic model.
We see from Eq.~(\ref{eq:G-loc}) and
Eqs.~(\ref{eq:tildeg1})~--~(\ref{eq:G-fin-A})
that the lowest order non local process 
are given by
\begin{equation}
\label{eq:proc1}
g^{\beta,\alpha} t^{\alpha,a} g^{a,a}_{1,1}
t^{a,\alpha} f^{\alpha,\alpha'}
t^{\alpha',a'} g^{a',a'}_{2,2} t^{a',\alpha'}
g^{\alpha',\beta}
\end{equation}
if the ferromagnets have an antiparallel spin orientation,
and by
\begin{eqnarray}
\label{eq:proc2}
g^{\beta,\alpha} t^{\alpha,a} g^{a,a}_{1,1}
t^{a,\alpha} g^{\alpha,\alpha'}
t^{\alpha',a'} g^{a',a'}_{1,1} t^{a',\alpha'}
f^{\alpha',\beta}\\
\label{eq:proc3}
f^{\beta,\alpha} t^{\alpha,a} g^{a,a}_{1,1}
t^{a,\alpha} g^{\alpha,\alpha'}
t^{\alpha',a'} g^{a',a'}_{1,1} t^{a',\alpha'}
g^{\alpha',\beta}
\end{eqnarray}
if the ferromagnets have a parallel spin orientation.
The corresponding diagrams are shown on
Fig.~\ref{fig:diag1}.
Because
each of these diagrams contains four
``$g$'' propagators and one ``$f$'' propagator,
the sign of the coefficient of
the non local term in $\Lambda^{\rm F}_{\rm Metal}$
and $\Lambda^{\rm AF}_{\rm Metal}$ is
positive (in agreement with Eqs.~\ref{eq:Lambda-AF}
and~\ref{eq:Lambda-F-fin}).
There is one diagram 
in the case of parallel spin orientations,
and there are two diagrams in the case of
antiparallel spin orientations, which
explains the factor of two
appearing in the case of a parallel
spin orientation~--~see
Eqs.~(\ref{eq:Lambda-AF})
and~(\ref{eq:Lambda-F-fin}).

Let us now consider insulating ferromagnets.
The lowest order diagrams are given
by~(\ref{eq:proc1}),~(\ref{eq:proc2})
and~(\ref{eq:proc3}) but now $g^{a,a}$
and $g^{a',a'}$ are real numbers.
As a consequence the sign of the non local
term with insulating ferromagnets
is opposite to the sign of the non local term
with metallic ferromagnets,
which is in agreement with Eqs.~(\ref{eq:Lambda-AF-In})
and~(\ref{eq:Lambda-F-In}).

\begin{figure}[thb]
\centerline{\fig{6cm}{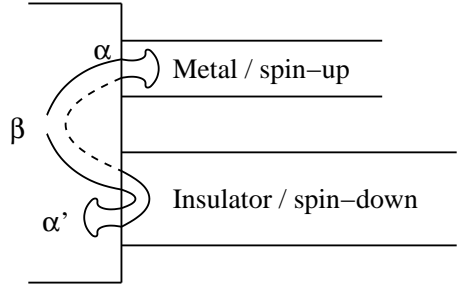}} 
\medskip
\caption{Lowest order processes in
the case of the mixed junction with antiparallel
spin orientations. } 
\label{fig:diag2}
\end{figure}

Finally in the mixed case the diagrams given
by~(\ref{eq:proc1}),~(\ref{eq:proc2}) and~(\ref{eq:proc3})
cancel because they are pure imaginary. Therefore we
look for the diagrams appearing in the next order.
One of these diagrams is
represented on Fig.~\ref{fig:diag2}. There are
four ``$g$'' propagators involved. The
sign of the diagram is positive, which explains
why the mixed junction behaves like the metallic
junction. The
diagram on Fig.~\ref{fig:diag2} is proportional
to $|t^{a,\alpha}|^2  |t^{a',\alpha'}|^4$, which is
in agreement with Eqs.~(\ref{eq:mixed1})
and~(\ref{eq:mixed2}).

\subsection{Ferromagnetic electrodes with
both spin channels}
\label{sec:h-ex}

In this section we calculate the superconducting gap of a
superconductor connected to one-dimensional ferromagnetic
electrodes having a partial spin polarization.
The motivation is to show that the results obtained
in the preceding section for half-metal
ferromagnets are valid also in the presence of a partial
spin polarization.
The ferromagnetic electrodes are described by the
Stoner Hamiltonian~(\ref{eq:H-Stoner}) in which
the exchange field is smaller than the Fermi energy.
The spin-up and spin-down channels are characterized
by the density of states $\rho_\uparrow$ and
$\rho_\downarrow$ (see Eqs.~(\ref{eq:rho-up})
and~(\ref{eq:rho-down})).
The derivations of the results
is given in Appendix~\ref{app:P}.

We use Eqs.~(\ref{eq:B4})~--~(\ref{eq:B7})
and Eqs.~(\ref{eq:B8})~--~(\ref{eq:B11})
obtained in Appendix~\ref{app:P}
and the local approximation already discussed
in section~\ref{sec:half-metal}. We find that
the high energy behavior of the Gorkov function
is given by~(\ref{eq:Gorkov-AF}) and~(\ref{eq:Gorkov-F}),
with the following parameters 
$\Lambda^{\rm Ferro}_{\rm Metal}$ and
$\Lambda^{\rm AF}_{\rm Metal}$:
\begin{eqnarray}
\label{eq:P-AntiFerro}
\LL \Lambda^{\rm AF}_{\rm Metal} \RR &=& 1
- \frac{1}{2}
\frac{a_0^2}{R_{\alpha,\beta}^2}
\frac{x_\uparrow + x_\downarrow}
{ (1+x_\uparrow)(1+x_\downarrow)}
- \frac{1}{2}
\frac{a_0^2}{R_{\alpha',\beta}^2}
\frac{x_\uparrow + x_\downarrow}
{ (1+x_\uparrow)(1+x_\downarrow)}\\
\nonb
&-& {1 \over 2}
\frac{a_0^3}{ R_{\alpha,\beta}
R_{\alpha',\beta} R_{\alpha,\alpha'}}
\frac{ x_\uparrow^2 + x_\downarrow^2
+2 x_\uparrow^2 x_\downarrow +
2 x_\uparrow x_\downarrow^2
+ 4 x_\uparrow x_\downarrow
- 2 x_\uparrow^2 x_\downarrow^2}
{(1+x_\uparrow)^2 (1+x_\downarrow)^2}\\
\label{eq:P-Ferro}
\LL \Lambda^{\rm Ferro}_{\rm Metal} \RR &=& 1
- \frac{1}{2}
\frac{a_0^2}{R_{\alpha,\beta}^2}
\frac{x_\uparrow + x_\downarrow}
{ (1+x_\uparrow)(1+x_\downarrow)}
- \frac{1}{2}
\frac{a_0^2}{R_{\alpha',\beta}^2}
\frac{x_\uparrow + x_\downarrow}
{ (1+x_\uparrow)(1+x_\downarrow)}\\
\nonb
&-& \frac{a_0^3}{ R_{\alpha,\beta}
R_{\alpha',\beta} R_{\alpha,\alpha'}}
\frac{ x_\uparrow^2 + x_\downarrow^2
+x_\uparrow^2 x_\downarrow +
x_\uparrow x_\downarrow^2
+ x_\uparrow x_\downarrow
- x_\uparrow^2 x_\downarrow^2}
{(1+x_\uparrow)^2 (1+x_\downarrow)^2}
,
\end{eqnarray}
where $x_\uparrow$ and
$x_\downarrow$ are given by
Eqs.~(\ref{eq:x-up}) and~(\ref{eq:x-down}).
The case of half-metal ferromagnets discussed
in section~\ref{sec:half-metal} can be recovered
by considering the limit $x_\downarrow=0$,
in which case Eqs.~(\ref{eq:P-AntiFerro})
and~(\ref{eq:P-Ferro}) reduce to Eqs.~(\ref{eq:HALF1})
and~(\ref{eq:HALF2}).
On the other hand it is easy to show from
Eqs.~(\ref{eq:P-AntiFerro}) and~(\ref{eq:P-Ferro})
that the ferromagnetic and antiferromagnetic
superconducting gaps are equal only if
there is no spin polarization
($\rho_\uparrow=\rho_\downarrow$).
As a consequence for two metallic ferromagnets
having an arbitrary small spin polarization,
the ferromagnetic gap is larger than the
antiferromagnetic gap.
This generalizes the behavior obtained in
section~\ref{sec:half-metal} in the case
of half-metal ferromagnets.

\section{Exact diagonalizations for half-metal ferromagnets}
\label{sec:num}
We present in this section a simulation based on
exact diagonalizations
in which  we can iterate the 
functional form of the
self consistency equation given by Eq.~(\ref{eq:self}).
The method is presented in
sections~\ref{sec:gen-BDG} and~\ref{sec:eval-green}.
The results are discussed in section~\ref{sec:theresults}.

\subsection{Bogoliubov-de Gennes
equations}
\label{sec:gen-BDG}
\subsubsection{Bogoliubov-de Gennes Hamiltonian}
\begin{figure}[thb]
\centerline{\fig{8cm}{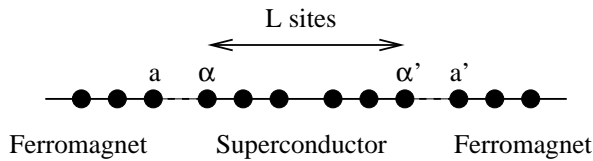}} 
\medskip
\caption{The geometry treated in the numerical
simulation. 
A superconductor on a one dimensional
segment with $L$ sites is connected to two
ferromagnets.  } 
\label{fig:geom}
\end{figure}

We consider the BCS model defined by
Eq.~(\ref{eq:tight}) on a one dimensional
lattice with $L$ sites (see Fig.~\ref{fig:geom}):
\begin{equation}
\label{eq:H-tight}
\hat{\cal H} -\mu \hat{N} = \sum_{\sigma,i=1}^L
-t \left( c_{i+1,\sigma}^+
c_{i,\sigma} + c_{i,\sigma}^+ c_{i+1,\sigma} \right)
+ \sum_{i=1}^L \Delta_i \left( c_{i,\uparrow}^+
c_{i,\downarrow}^+ + c_{i,\downarrow}
c_{i,\uparrow} \right) - \mu \hat{N}
.
\end{equation}
The two dimensional model cannot be treated numerically
because of computational limitations and this is why
we consider a one dimensional geometry.
Nevertheless, the method that we use in 1D can be
also applied to 2D models.
It is convenient to use the notation
\begin{equation}
\label{eq:check-psi}
\check{\psi}_\uparrow^+ =
\left[ c_{1,\uparrow}^+ , ..., c_{L,\uparrow}^+,
c_{1,\downarrow},...,c_{L,\downarrow} \right]
,
\end{equation}
in which $\check{\psi}_\uparrow^+$ has $2L$ components.
We use~(\ref{eq:check-psi}) to obtain the
Bogoliubov-de Gennes Hamiltonian
\begin{equation}
\label{eq:H-2Nx2N}
\check{H} = \check{\psi}_\uparrow^+
\check{K} \check{\psi}_\uparrow+
\check{\psi}_\uparrow^+
\check{\Delta}\check{\psi}_\uparrow
,
\end{equation}
where the kinetic term is
\begin{eqnarray}
\check{K}^{1,1}_{i,j} &=& - t \left(
\delta_{i,j+1} + \delta_{i,j-1} \right)
+ \mu \delta_{i,j} \\
\check{K}^{2,2}_{k,l} &=&  t \left(
\delta_{k,l+1} + \delta_{k,l-1} \right)
- \mu \delta_{k,l} \\
\check{K}^{1,2}_{i,k} &=& \check{K}^{2,1}_{k,i} =0
,
\end{eqnarray}
and the pairing term is
\begin{eqnarray}
\check{\Delta}^{1,1}_{i,j} &=& \check{\Delta}^{2,2}_{i,j} =0\\
\check{\Delta}^{1,2}_{i,k} &=& \check{\Delta}^{2,1}_{k,i} =
\Delta_i \delta_{i,k}
.
\end{eqnarray}
Similarly to the Nambu representation, we have
used the label ``1'' for the
``electronic'' components of $\check{\psi}$ and the
label ``2'' for the``hole'' components.
We have doubled the space coordinates:
the labels $i,j$ correspond to the electronic component
and the labels $k,l$ correspond to the hole component.
The symbol $\delta_{i,k}$ means that
$i$ and $k$ correspond to the same site on the lattice but
belong to a different Nambu component.

\subsubsection{Spectral representations}
\label{sec:spectral-rep}
The eigenvectors of the Bogoliubov-de Gennes Hamiltonian
(\ref{eq:H-2Nx2N}) take the form
\begin{eqnarray}
\label{eq:psi-alpha}
|\psi_\alpha \rangle &=& \sum_{i=1}^L
R_{\alpha,i} | e_i \rangle
+ \sum_{k=1}^L R_{\alpha,k} |e_k \rangle \\
|\psi_\beta \rangle &=& \sum_{i=1}^L
R_{\beta,i} | e_i \rangle
+ \sum_{k=1}^L R_{\beta,k} |e_k \rangle
\label{eq:psi-beta}
,
\end{eqnarray}
where the eigenvalues are such that $\lambda_\alpha>0$
and $\lambda_\beta<0$. In this notation there are
$L$ kets $|e_i\rangle$ associated to the first
component of the Nambu representation, and
there are $L$ kets $|e_k \rangle$ associated
to the second component of the Nambu representation.
We deduce from Eqs. (\ref{eq:psi-alpha}) and (\ref{eq:psi-beta})
the form of the quasiparticle operators
\begin{eqnarray}
\Gamma_{\alpha,\downarrow}^+ &=&
\sum_i R_{\alpha,i} c_{i,\uparrow}
+ \sum_k R_{\alpha,k} c_{k,\downarrow}^+ \\
\Gamma_{\beta,\uparrow} &=&
\sum_i R_{\beta,i} c_{i,\uparrow}
+ \sum_k R_{\beta,k} c_{k,\downarrow}^+ 
\end{eqnarray}
which diagonalize the Bogoliubov-de Gennes Hamiltonian
$$
\hat{\cal H} = \sum_\alpha \lambda_\alpha
\Gamma_{\alpha,\downarrow}^+
\Gamma_{\alpha,\downarrow} -
\sum_\beta \lambda_\beta
\Gamma_{\beta,\uparrow}^+ \Gamma_{\beta,\uparrow}
.
$$
The spectral representation of the
Green's function~(\ref{eq:Green-def}) can be
expressed in terms of the matrix $R$:
\begin{eqnarray}
\label{eq:RS-1}
g_{i,j}^{A,1,1} &=& \sum_\beta \frac{ R_{\beta,i}
R_{\beta,j}} {\omega + i \eta - \left[ \mu
+ |E_\beta| \right]} +
\sum_\alpha \frac{ R_{\alpha,i}
R_{\alpha,j}} {\omega + i \eta - \left[ \mu
- E_\alpha \right]}\\
g_{i,k}^{A,1,2} &=& \sum_\beta \frac{ R_{\beta,i}
R_{\beta,k}} {\omega + i \eta - \left[ \mu
+ |E_\beta| \right]} +
\sum_\alpha \frac{ R_{\alpha,i}
R_{\alpha,k}} {\omega + i \eta - \left[ \mu
- E_\alpha \right]}
\label{eq:RS-2}
.
\end{eqnarray}

\subsection{Evaluation of the Green's functions}
\label{sec:eval-green}
\subsubsection{Evaluation of a spectral representation:}
The Green's functions are obtained from
Eqs.~(\ref{eq:RS-1}) and~(\ref{eq:RS-2}) in terms of
their poles $\omega_n$ and residues $R_n$:
\begin{equation}
\label{eq:f}
\label{eq:g-res}
g_0^A(\omega) = \sum_n \frac{R_n}{\omega - \omega_n - i \eta}
.
\end{equation}
To make the integration over
energy, we go to the limit of zero dissipation ($\eta \rightarrow 0$)
and use the identity $1/[\omega - \omega_n - i \eta]
= {\cal P}/[\omega - \omega_n] + i \pi \delta(\omega
- \omega_n)$. 
To show that the principal part can be neglected
if $\omega>\Delta$, we
come back to the particular case where the
superconducting order parameter is uniform:
$\Delta_\beta \equiv \Delta_0$ for all $\beta$.
In this case the spectral representation takes
the form
\begin{eqnarray}
\label{eq:spec-supra-1}
g_{\alpha,\beta}^{A}
(\omega) &=&\frac{1}{\cal N}
\sum_\vk e^{i\vk.(\vx_\alpha-\vx_\beta)}
\left[
\frac{ |u_k|^2}{\omega - (\mu_S+E_k)-i\eta} +
\frac{ |v_k|^2}{\omega - (\mu_S-E_k)-i\eta}
\right]
.
\end{eqnarray}
We start from Eq.~(\ref{eq:spec-supra-1}) and make the
substitution
\begin{eqnarray}
\frac{1}{\omega-(\mu_S+E_k) - i \eta}
\rightarrow i \pi \delta \left(\omega-(\mu_S+E_k) \right)\\
\frac{1}{\omega-(\mu_S-E_k) - i \eta}
\rightarrow i \pi \delta \left(\omega-(\mu_S-E_k) \right)
.
\end{eqnarray}
The Green's function given by Eq.~(\ref{eq:spec-supra-1})
becomes
\begin{equation}
\label{eq:delta-f}
g_{\alpha,\beta}^{A} \rightarrow
i \pi {1 \over {\cal N}} \sum_\vk
e^{i\vk.(\vx_\alpha-\vx_\beta)}
\left[ (u_k)^2 \delta \left(\omega-(\mu_S+E_k)\right)
+ (v_k)^2 \delta \left(\omega-(\mu_S-E_k)\right) \right]
.
\end{equation}
After using the $\delta$-function~(\ref{eq:delta-f})
and performing the integral over wave vector we recover
the form~(\ref{eq:Green}) of the Green's function
in which the term proportional to $\cos{ \varphi}$
has been discarded. The fact that the term
proportional to $\cos{\varphi}$ is not included
does not constitute a
problem because we know from section~\ref{sec:local}
that the envelope of the $2 k_F$-oscillations
is the same in the presence or absence of
the $\cos{\varphi}$ term.
Therefore if $\omega>\Delta$,
Eq.~(\ref{eq:g-res}) can be replaced by
\begin{equation}
\label{eq:g0-delta}
g_0^A(\omega)=i\pi \sum_n R_n \delta \left(
\omega - \omega_n \right)
.
\end{equation}

\subsubsection{Evaluation of the $\delta$-functions}
\label{sec:delta}
To evaluate the $\delta$-function in (\ref{eq:g0-delta}),
we replace
$\delta (\omega - \omega_n)$ by $\delta_\eta(\omega)$,
where $\delta_\eta(\omega)$ is a function having a width
$\eta$ in energy, and normalized to unity:
$\int \delta_\eta(\omega) d \omega = 1$. For instance
$\delta_\eta(\omega)$ can be chosen as a Lorentzian or a Gaussian.
To obtain the value of a Green's function at a single energy
$\omega$ the Lorentzian or Gaussian will be
evaluated $2 L$ times. 
To optimize this part of the program, it is useful to use
a function $\delta_\eta(\omega)$ that is finite only in
the interval $\left[ - \eta, \eta \right]$ and
vanishes outside this energy interval. The simplest choice
is given by
\begin{equation}
\label{eq:para}
\delta_\eta(\omega) =
{3 \over 4 \eta} \left[ 1 - \left(
{\omega \over \eta} \right)^2 \right]
\mbox{ if $|\omega|<\eta$. }
\end{equation}

\subsection{Results}
\label{sec:theresults}
We consider the geometry represented on Fig.~\ref{fig:geom}
in which a one dimensional superconductor
on an open segment with $L$ sites is connected
to two ferromagnetic metals.
The superconductor is described
by the BCS tight-binding Hamiltonian~(\ref{eq:tight}).
We note $t_0=t_{a,\alpha}/t=t_{a',\alpha'}/t$ the tunnel matrix element
connecting the ferromagnets and the superconductor,
normalized to the bandwidth of the superconductor.
Low transparency interfaces correspond to
$t_0 \ll 1$ and high transparency interfaces
correspond to $t_0 \sim 1$.

\subsubsection{Density of states}
\begin{figure}[thb]
\centerline{\fig{10cm}{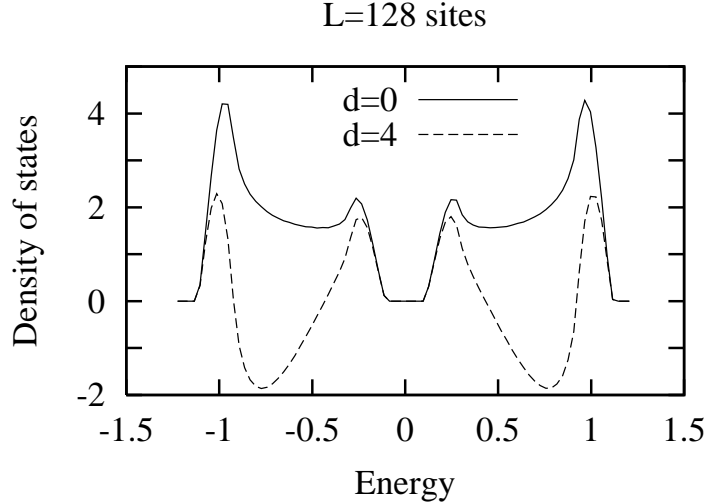}} 
\medskip
\caption{Energy dependence of the density
of state $\rho_g^{\alpha,\beta}$, for two
values of the distance between the sites
$\alpha$ and $\beta$. We used
periodic boundary conditions, with
$L=128$ sites. The hopping energy is
$t=0.5$, the superconducting
gap $\Delta_0=0.2$ is uniform and
the level broadening is $\eta=0.1$. } 
\label{fig:DOS128}
\end{figure}
We have shown on Fig.~\ref{fig:DOS128}
the energy dependence of the density of states
$\rho_g^{\alpha,\beta}$ associated to the
ordinary propagator (see section~\ref{sec:themethod})
in the presence of a uniform
gap profile $\Delta_\beta\equiv\Delta_0$ for all $\beta$,
and with $L=128$ sites.
It is visible on this figure that 
the different parameters
of the simulation are compatible with each other.
Namely, the level broadening $\eta$
is sufficiently small to have a well-defined
superconducting gap.
The level broadening is also sufficiently large
for quasiparticle states to form a continuous
band. Because of these two constraints,
we cannot use in this simulation
realistic parameters as we did 
for the local approach in section~\ref{sec:local}
(see Fig.~\ref{fig:gap-1chan}). Using realistic
parameters would require too large system
sizes.

Finally, the calculation of the superconducting order parameter
presented in section~\ref{sec:local}
were based on the estimation of
the Gorkov function at high energy.
By contrast low energy degrees of freedom play a relevant role
in our simulation.
One of the questions
that will be answered by the exact diagonalizations
is to determine whether
low energy degrees of freedom (probed by the numerical simulation
with strong finite size effects)
have the same physics as
high energy degrees of freedom (probed by the local 
approach in section~\ref{sec:local}).

\subsubsection{Gap profile}
\begin{figure}[thb]
\centerline{\fig{10cm}{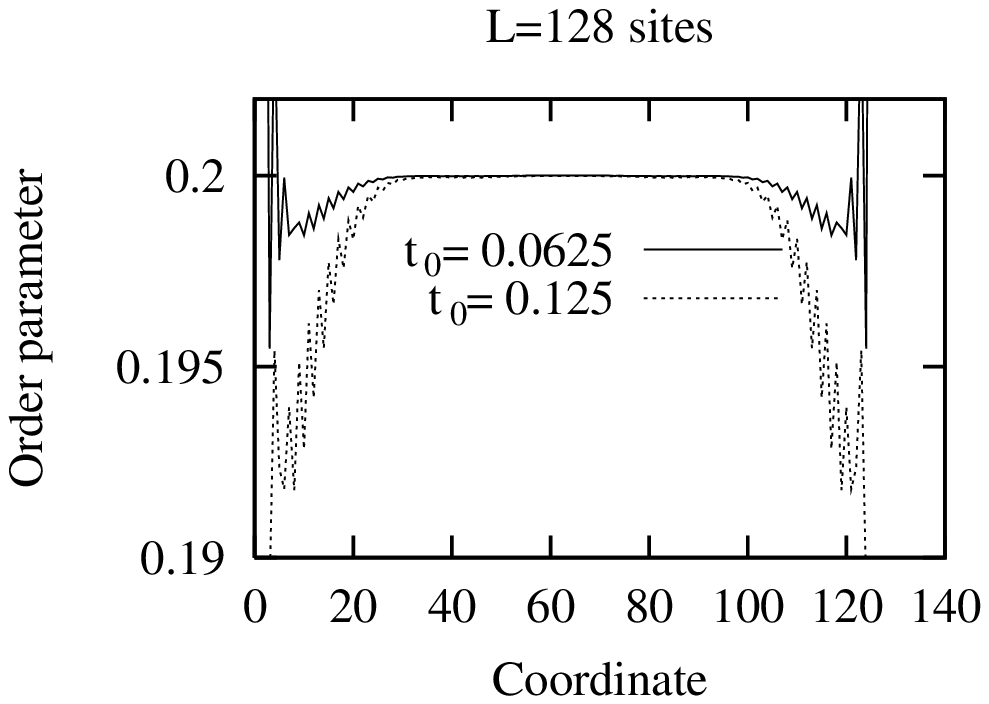}} 
\medskip
\caption{ Self consistent gap profile with
$L=128$ sites and
two values of $t_0$:
$t_0=0.0625$ (solid line) and $t_0=0.125$ (dotted
line). The other parameters are the same as
on Fig.~\ref{fig:DOS128}. 
The difference between the parallel and the
antiparallel superconducting order parameters cannot be
distinguished on the
scale of the figure. } 
\label{fig:prof128}
\end{figure}
The gap profile is shown on
Fig.~\ref{fig:prof128}
for $L=128$ sites. We have obtained similar results for
$L=32$ and $L=64$ sites. The gap profile obtained
with exact diagonalizations is qualitatively
similar to
section~\ref{sec:local}. Namely, the superconducting
order parameter
is reduced close to the interface with the ferromagnets and
we obtain $2 k_F$ oscillations in the gap profile.

\subsubsection{Difference between the superconducting gaps
in the ferromagnetic and antiferromagnetic alignments}
\begin{figure}[thb]
\centerline{\fig{10cm}{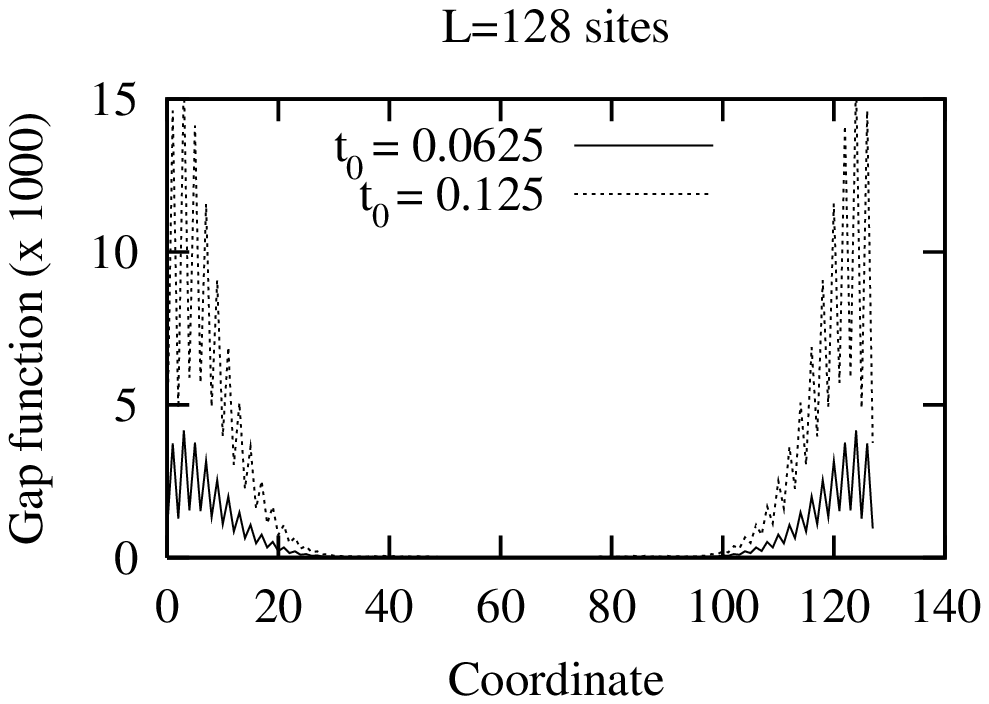}} 
\medskip
\caption{Variation of $\delta_\beta$ defined
by~(\ref{eq:gap-ratio-def})
with $L=128$ sites in the superconductor.
The parameters are identical to
Fig.~\ref{fig:prof128}. } 
\label{fig:ratio128}
\end{figure}
We have shown on Fig.~\ref{fig:ratio128} the
variation of $\delta_\beta$ defined
by Eq.~(\ref{eq:gap-ratio-def})
with $L=128$ sites.
Similar results have been obtained with
$L=32$ and  $L=64$ sites.
For each site $\beta$ in the superconductor,
we have calculated the superconducting order parameters
$\Delta^F_\beta$ and $\Delta^{\rm AF}_\beta$ with
parallel and antiparallel
spin orientations in the two ferromagnetic electrodes.
From what we
deduce the value of $\delta_\beta$ defined
by Eq.~(\ref{eq:gap-ratio-def}). The calculation
consists in iterating the process
on Fig.~\ref{fig:process} until a sufficient precision
has been obtained. The relative
error made in the determination of the order 
parameters is several orders of magnitude
smaller than the difference between the
ferromagnetic and antiferromagnetic
superconducting gaps.

We made the simulations with two
values of the normalized hopping between
the superconductor and the
ferromagnet (see Fig.~\ref{fig:ratio128}).
We also tried larger
values of the interface transparencies but
the algorithm did not converge.
The clarification of this point is left as an
open question for future work.
From the result presented on
Fig.~\ref{fig:ratio128} we deduce that
\begin{itemize}
\item[(i)] With all available sizes and
interface transparencies, $\delta_\beta$
defined by (\ref{eq:gap-ratio-def})
is positive, meaning that {\sl the gap is
larger in the ferromagnetic alignment}.
This is opposite to the model with insulating
ferromagnets~\cite{deGennes} and is in
agreement with the approaches used
in sections~\ref{sec:half-metal} and~\ref{sec:h-ex}.
\item[(ii)] $\delta_\beta$ defined
by Eq.~(\ref{eq:gap-ratio-def})
tends to zero in the bulk of the
superconductor. The cross-over between the
surface and bulk behavior is controlled
by a length scale which is of order 10
on Fig.~\ref{fig:ratio128}.
It is expected that 
this length scale is equal to
the superconducting coherence length
given by
$\xi_0=\epsilon_F / (k_F \Delta_0)$.
\end{itemize}

\section{Conclusion}
\label{sec:conclu}
To summarize, we have provided a detailed
investigation of F/S/F trilayers with
metallic ferromagnets.
We found that
the physics of the metallic trilayer
was dominated by pair correlations,
not by pair breaking. 
This behavior was obtained with several 
complementary approaches:
\begin{itemize}
\item[(i)] An approach based on the estimation
of the high energy behavior of the Gorkov function
for  half-metal ferromagnets (see section~\ref{sec:local})
and for ferromagnets having both spin conduction channels
(see section~\ref{sec:h-ex}).
In these approaches we could
use realistic parameters
($\Delta_{\rm bulk}/D = 10^{-4}$).
\item[(ii)] {\sl Exact diagonalizations}
(see section~\ref{sec:num}) that were limited
to small sizes and large values of
$\Delta_{\rm bulk}/D$ ($\Delta_{\rm bulk}/D =0.2 $).
\end{itemize}
In all approaches we find that the F/S/F trilayer
with metallic ferromagnets is characterized
by $\Delta_{\rm F} > \Delta_{\rm AF}$ while
the F/S/F trilayer with insulating ferromagnets
is characterized by $\Delta_{\rm AF} > \Delta_{\rm F}$.

Finally, we
mention two recent theoretical
articles~\cite{Buzdin}
in which Usadel equations have been used to treat
F/S/F heterostructures in the diffusive regime.
These authors
have found that the metallic F/S/F heterostructure
was controlled by pair breaking ($\Delta_{\rm F}
< \Delta_{\rm AF}$) while we have found here an
opposite behavior ($\Delta_{\rm F} > \Delta_{\rm AF}$).
In fact we believe that both approaches are correct
but do not incorporate the same ingredients.
The behavior of the model that
we consider here is strongly related to non local
pair correlations and can be explained with
lowest order diagrams. The existence of a simple
explanation shows the validity of the picture proposed
in our article. On the other hand the model
considered in Ref.~\cite{Buzdin} is well suited
to describe diffusive heterostructures.
We are not certain at the present stage that
the usual form of Usadel equations can be
used to describe non local processes as
we did here in ballistic systems.
We think that a lot of understanding can be
gained by discussing non local Usadel
equations.

\section*{Acknowledgments}
The authors acknowledge fruitful discussions
with A. Buzdin, J.C. Cuevas and D. Feinberg.

\appendix

\section{Expression of the Keldysh propagators}
\label{app:Keldysh}
The goal of this appendix is to rederive the
main results of this article with the
non equilibrium
form~(\ref{eq:Dy2}) of the Gorkov function,
rather than the equilibrium Gorkov function~(\ref{eq:Gpm-eq}).
This formalism based on non equilibrium Green's
functions is more general than the equilibrium
Green's function formalism because it can
also be applied to non equilibrium problems.
A detailed investigation of this issue will be
presented in the future.
Here, we want to show that both formalisms
coincide for the equilibrium problem, which
constitutes also a test of the calculations
presented in the main body of the article.

\subsection{One-channel problem}
\label{sec:one-channel}

Let us first consider the  single
channel model (see Fig.~\ref{fig:1channel}).
The Green's functions $G^{R,A}$ are the solution of
the Dyson equation
$$
\HG^{a,a} = \Hg^{a,a} +
\Hg^{a,a} \Ht^{a,\alpha} \Hg^{\alpha,\alpha}
\Ht^{\alpha,a}
\HG^{a,a}
,
$$
from what we deduce
$G^{a,a}_{1,1} = g^{a,a}_{1,1} /{\cal D}$, with
${\cal D}$ given by~(\ref{eq:D-1contact}).
The Dyson-Keldysh equation
associated to an arbitrary site in the superconductor
takes the form
\begin{equation}
\HG^{+,-}_{\beta,\beta} = \Hg^{+,-}_{\beta,\beta}
+ \Hg^{+,-}_{\beta,\alpha} \Ht_{\alpha,a}
\HG^A_{a,\beta} 
+ \HG^R_{\beta,a} \Ht_{\alpha,a} \Hg^{+,-}_{\alpha,\beta}
+ \HG^R_{\beta,a} \Ht_{a,\alpha} \Hg^{+,-}_{\alpha,\alpha}
\Ht_{\alpha,a} \HG^A_{a,\beta}
+ \HG^R_{\beta,\alpha} \Ht_{\alpha,a}
\Hg^{+,-}_{a,a} \Ht_{a,\alpha}
\HG^A_{\alpha,\beta}
\label{eq:G-beta1}
.\end{equation}
Evaluating the five terms in (\ref{eq:G-beta1})
leads to
\begin{eqnarray}
G^{+,-}_{\beta,\beta} &=&
2 i \pi n_F(\omega-\mu_S) \left\{ \rho_{f}^{\beta,\beta}
+ |t_{a,\alpha}|^2 {1 \over {\cal D}^A}
g^{a,a,A}_{1,1} f^{\alpha,\beta,A} \rho_g^{\beta,\alpha}
+ |t_{a,\alpha}|^2 {1 \over {\cal D}^R}
g^{a,a,R}_{1,1} g^{\beta,\alpha,R} \rho_f^{\alpha,\beta}
\right.\\
\nonb
&+& \left.|t^{a,\alpha}|^4 \frac{1}{ {\cal D}^A {\cal D}^R}
g^{a,a,A}_{1,1} g^{a,a,R}_{1,1} g^{\beta,\alpha,R}
f^{\alpha,\beta,A} \rho_g^{\alpha,\alpha} \right\}
+ 2 i \pi n_F(\omega-\mu_a) |t^{a,\alpha}|^2
\frac{1}{ {\cal D}^A {\cal D}^R }
g^{\beta,\alpha,R} f^{\alpha,\beta,A}
\rho^{a,a}_{1,1}
\nonb
.
\end{eqnarray}
The final step is to show that with $\mu_a = \mu_S$
this expression coincides with~(\ref{eq:Gorkov-local-1ch}).

\subsection{Two-channel problem with antiparallel magnetizations}
\label{sec:antipara}

The Dyson-Keldysh equation associated to an arbitrary site
$\beta$ in the superconductor is the following:
\begin{eqnarray}
\label{eq:Gpm-gene}
\HG^{+,-}_{\beta,\beta} &=& \Hg^{+,-}_{\beta,\beta}
+ \Hg^{+,-}_{\beta,\alpha} \Ht_{\alpha,a}
\HG^A_{a,\beta} + \Hg^{+,-}_{\beta,\alpha'}
\Ht_{\alpha',a'} \HG^A_{a',\beta}
+ \HG^R_{\beta,a} \Ht_{a,\alpha} \Hg^{+,-}_{\alpha,\beta}
+ \HG^R_{\beta,a'} \Ht_{a',\alpha'} \Hg^{+,-}_{\alpha',\beta}
+ \HG^R_{\beta,a} \Ht_{a,\alpha} \Hg^{+,-}_{\alpha,\alpha}
\Ht_{\alpha,a} \HG^A_{a,\beta}\\
\nonb
&+& \HG^R_{\beta,a'} \Ht_{a',\alpha'} \Hg^{+,-}_{\alpha',\alpha'}
\Ht_{\alpha',a'} \HG^A_{a',\beta}
+ \HG^R_{\beta,a} \Ht_{a,\alpha} \Hg^{+,-}_{\alpha,\alpha'}
\Ht_{\alpha',a'} \HG^A_{a',\beta}
+ \HG^R_{\beta,a'} \Ht_{a',\alpha'} \Hg^{+,-}_{\alpha',\alpha}
\Ht_{\alpha,a} \HG^A_{a,\beta}\\\nonb
&+& \HG^R_{\beta,\alpha} \Ht_{\alpha,a} \Hg^{+,-}_{a,a}
\Ht_{a,\alpha} \HG^A_{\alpha,\beta}
+ \HG^R_{\beta,\alpha'} \Ht_{\alpha',a'} \Hg^{+,-}_{a',a'}
\Ht_{a',\alpha'} \HG^A_{\alpha',\beta}
.
\end{eqnarray}

We need the expression of the following
Green's functions:
\begin{eqnarray}
\label{eq:G-debut}
\HG^{\beta,a} &=& t^{a,\alpha} g^{a,a}_{1,1}
\left[\begin{array}{cc}
\tilde{g}^{\beta,\alpha} & 0\\
\tilde{f}^{\beta.\alpha} & 0 \\
\end{array} \right]\\
\HG^{\beta,a'} &=& -t^{a',\alpha'} g^{a',a'}_{2,2}
\left[\begin{array}{cc}
0 & \tilde{f}^{\beta,\alpha'} \\
0 & \tilde{g}^{\beta,\alpha'} \\
\end{array} \right]\\
\HG^{\alpha,\beta} &=& \left[ \begin{array}{cc}
\tilde{g}^{\alpha,\beta} & \tilde{f}^{\alpha,\beta} \\
G^{\alpha,\beta}_{2,1} & G^{\alpha,\beta}_{2,2} 
\end{array} \right] \\
\HG^{\beta,\alpha} &=& \left[ \begin{array}{cc}
\tilde{g}^{\beta,\alpha} & \tilde{G}^{\beta,\alpha}_{1,2} \\
\tilde{f}^{\beta,\alpha} & G^{\beta,\alpha}_{2,2} 
\end{array} \right] \\
\HG^{\alpha',\beta} &=& \left[ \begin{array}{cc}
G^{\alpha',\beta}_{1,1} & G^{\alpha',\beta}_{1,2} \\
\tilde{f}^{\alpha',\beta} & \tilde{g}^{\alpha',\beta} 
\end{array} \right] \\
\HG^{\beta,\alpha'} &=& \left[ \begin{array}{cc}
G^{\beta,\alpha'}_{1,1} & \tilde{f}^{\beta,\alpha'} \\
G^{\beta,\alpha'}_{2,1} & \tilde{g}^{\beta,\alpha'}
\label{eq:G-fin}
\end{array} \right] 
.
\end{eqnarray}
We deduce from~(\ref{eq:G-debut-A})~--~(\ref{eq:G-fin-A})
and~(\ref{eq:G-debut})~--~(\ref{eq:G-fin})
the final form of the Gorkov function in the antiparallel
alignment:
\begin{eqnarray}
\label{eq:local-Gorkov-AF}
\HG^{+,-}_{\beta,\beta} &=&
2 i \pi n_F(\omega-\mu_S) \left\{
\rho_{f}^{\beta,\beta} +
|t^{a,\alpha}|^2 g^{a,a,A}_{1,1} \tilde{f}^{\alpha,\beta,A}
\rho_g^{\beta,\alpha}
+ |t^{a',\alpha'}|^2 g^{a',a',A}_{2,2} \tilde{g}^{\alpha',\beta,A}
\rho_f^{\beta,\alpha'} 
+ |t^{a,\alpha}|^2 g^{a,a,R}_{1,1}
\tilde{g}^{\beta,\alpha,R} \rho_f^{\alpha,\beta}
\right. \\
\nonb
&+& |t^{a',\alpha'}|^2 g^{a',a',R}_{2,2}
\tilde{f}^{\beta,\alpha',R} \rho_g^{\alpha',\beta}
+ |t^{a,\alpha}|^4 g^{a,a,R}_{1,1} g^{a,a,A}_{1,1}
\tilde{g}^{\beta,\alpha,R} \tilde{f}^{\alpha,\beta,A}
\rho_g^{\alpha,\alpha}
+ |t^{a',\alpha'}|^4 g^{a',a',R}_{2,2} g^{a',a',A}_{2,2}
\tilde{g}^{\alpha',\beta,A} \tilde{f}^{\beta,\alpha',R}
\rho_g^{\alpha',\alpha'}\\
\nonb
&+& \left. |t^{a,\alpha}|^2 |t^{a',\alpha'}|^2
g^{a,a,R}_{1,1} g^{a',a',A}_{2,2}
\tilde{g}^{\beta,\alpha,R} \tilde{g}^{\alpha',\beta,A}
\rho_f^{\alpha,\alpha'}
+ |t^{a,\alpha}|^2 |t^{a',\alpha'}|^2
g^{a,a,A}_{1,1} g^{a',a',R}_{2,2}
\tilde{f}^{\beta,\alpha',R} \tilde{f}^{\alpha,\beta,A}
\rho_f^{\alpha',\alpha} \right\}\\
&+& 2 i \pi n_F(\omega-\mu_a) |t^{a,\alpha}|^2
\tilde{g}^{\beta,\alpha,R} \tilde{f}^{\alpha,\beta,A}
\rho_{1,1}^{a,a}
+ 2 i \pi n_F(\omega-\mu_{a'})
|t^{a',\alpha'}|^2
\tilde{f}^{\beta,\alpha',R}
\tilde{g}^{\alpha',\beta,A} \rho_{2,2}^{a',a'}
.
\nonb
\end{eqnarray}
Using the propagators obtained in
section~\ref{sec:main-AF} to
evaluate~(\ref{eq:local-Gorkov-AF}), we can show that
Eq.~(\ref{eq:local-Gorkov-AF}) leads
directly to~(\ref{eq:Gorkov-AF}) and~(\ref{eq:Lambda-AF}).

\subsection{Two-channel problem with parallel magnetizations}
Let us now
consider two single-channel ferromagnetic electrodes having
a parallel spin orientation. 
Following section~\ref{sec:antipara}, we obtain
\begin{eqnarray}
\label{eq:GORKOV-P}
\HG^{+,-}_{\beta,\beta} &=&
2 i \pi n_F(\omega-\mu_S) \left\{
\rho_{f}^{\beta,\beta} +
|t^{a,\alpha}|^2 g^{a,a,A}_{1,1} \tilde{f}^{\alpha,\beta,A}
\rho_g^{\beta,\alpha}
+ |t^{a',\alpha'}|^2 g^{a',a',A}_{1,1} \tilde{f}^{\alpha',\beta,A}
\rho_g^{\beta,\alpha'} 
+ |t^{a,\alpha}|^2 g^{a,a,R}_{1,1}
\tilde{g}^{\beta,\alpha,R} \rho_f^{\alpha,\beta} \right.\\
\nonb
&+& |t^{a',\alpha'}|^2 g^{a',a',R}_{1,1}
\tilde{g}^{\beta,\alpha',R} \rho_f^{\alpha',\beta}
+ |t^{a,\alpha}|^4 g^{a,a,R}_{1,1} g^{a,a,A}_{1,1}
\tilde{g}^{\beta,\alpha,R} \tilde{f}^{\alpha,\beta,A}
\rho_g^{\alpha,\alpha}
+ |t^{a',\alpha'}|^4 g^{a',a',R}_{1,1} g^{a',a',A}_{1,1}
\tilde{f}^{\alpha',\beta,A} \tilde{g}^{\beta,\alpha',R}
\rho_g^{\alpha',\alpha'}\\
\nonb
&+& \left. |t^{a,\alpha}|^2 |t^{a',\alpha'}|^2
g^{a,a,R}_{1,1} g^{a',a',A}_{1,1}
\tilde{g}^{\beta,\alpha,R} \tilde{f}^{\alpha',\beta,A}
\rho_g^{\alpha',\alpha}
+ |t^{a,\alpha}|^2 |t^{a',\alpha'}|^2
g^{a,a,A}_{1,1} g^{a',a',R}_{1,1}
\tilde{g}^{\beta,\alpha',R} \tilde{f}^{\alpha,\beta,A}
\rho_g^{\alpha',\alpha} \right\}\\
&+& 2 i \pi n_F(\omega-\mu_a) |t^{a,\alpha}|^2
\tilde{g}^{\beta,\alpha,R} \tilde{f}^{\alpha,\beta,A}
\rho_{1,1}^{a,a}
+ 2 i \pi n_F(\omega-\mu_{a'})
|t^{a',\alpha'}|^2
\tilde{g}^{\beta,\alpha',R}
\tilde{f}^{\alpha',\beta,A} \rho_{1,1}^{a',a'}
,
\nonb
\end{eqnarray}
where the propagators $\tilde{g}$ and $\tilde{f}$ are given by
\begin{eqnarray}
\tilde{g}^{\alpha,\beta} &=& \frac{1}{{\cal D}_{\rm F}}
\left[ g^{\alpha,\beta} + |t^{a',\alpha'}|^2
g^{a',a'}_{2,2} \left( g^{\alpha,\alpha'}
g^{\alpha'\beta} - g^{\alpha',\alpha'}
g^{\alpha,\beta} \right) \right]\\
\tilde{f}^{\alpha,\beta} &=& \frac{1}{{\cal D}_{\rm F}}
\left[ f^{\alpha,\beta} + |t^{a',\alpha'}|^2
g^{a',a'}_{2,2} \left( g^{\alpha,\alpha'}
f^{\alpha',\beta} - g^{\alpha',\alpha'}
f^{\alpha,\beta} \right) \right]\\
\tilde{g}^{\beta,\alpha} &=& \frac{1}{{\cal D}_{\rm F}}
\left[ g^{\beta,\alpha} + |t^{a',\alpha'}|^2
g^{a',a'}_{2,2} \left( g^{\alpha',\alpha}
g^{\beta,\alpha'} - g^{\alpha',\alpha'}
g^{\beta,\alpha} \right) \right]\\
\tilde{f}^{\beta,\alpha} &=& \frac{1}{{\cal D}_{\rm F}}
\left[ f^{\beta,\alpha} + |t^{a',\alpha'}|^2
g^{a',a'}_{2,2} \left( g^{\alpha',\alpha}
f^{\beta,\alpha'} - g^{\alpha',\alpha'}
f^{\beta,\alpha} \right) \right]
,
\end{eqnarray}
where ${\cal D}_F$ is given by Eq.~(\ref{eq:D-F}).
We can show that
Eq.~(\ref{eq:GORKOV-P}) leads directly to~(\ref{eq:Gorkov-F})
and~(\ref{eq:Lambda-F-fin}).

\section{Ferromagnetic electrodes with an
arbitrary spin polarization}
\label{app:P}

\subsection{Dyson matrix}

The advanced and retarded Green's function $\HG_{\beta,\beta}$
at an arbitrary site $\beta$ of the superconductor
can be deduced from Eq.~(\ref{eq:Dy1}):
\begin{equation}
\label{eq:G-beta-beta}
G^{\beta,\beta}_{2,1} =
g^{\beta,\beta}_{2,1} +
g^{\beta,\alpha}_{2,1} t^{\alpha,a} G^{a,\beta}_{1,1}
- g^{\beta,\alpha}_{2,2} t^{\alpha,a} G^{a,\beta}_{2,1} 
+ g^{\beta,\alpha'}_{2,1} t^{\alpha',a'} G^{a',\beta}_{1,1}
- g^{\beta,\alpha'}_{2,2} t^{\alpha',a'} G^{a',\beta}_{2,1}
.
\end{equation}
Eq.~(\ref{eq:G-beta-beta}) can be used to evaluate
the equilibrium Gorkov function given by Eq.~(\ref{eq:Gpm-eq})
and deduce the value of the self-consistent superconducting
order parameter. The Green's functions
$G^{a,\beta}_{1,1}$, $G^{a,\beta}_{2,1}$,
$G^{a',\beta}_{1,1}$ and $G^{a',\beta}_{2,1}$ are the
solution of the Dyson equation~(\ref{eq:Dy1}) which can
be expressed as a $4 \times 4$ Dyson matrix:
\begin{equation}
\label{eq:Dy-4x4}
\left[\begin{array}{cccc}
1 - K_{1,1}^{a,\alpha} t^{\alpha,a} &
K_{1,2}^{a,\alpha} t^{\alpha,a} &
- K_{1,1}^{a,\beta} t^{\beta,b} &
K_{1,2}^{a,\beta} t^{\beta,b} \\
K_{2,1}^{a,\alpha} t^{\alpha,a} &
1 - K_{2,2}^{a,\alpha} t^{\alpha,a} &
K_{2,1}^{a,\beta} t^{\beta,b} &
- K_{2,2}^{a,\beta} t^{\beta,b} \\
-K_{1,1}^{b,\alpha} t^{\alpha,a} &
K_{1,2}^{b,\alpha} t^{\alpha,a} &
1-K_{1,1}^{b,\beta} t^{\beta,b} &
K_{1,2}^{b,\beta} t^{\beta,b} \\
K_{2,1}^{b,\alpha} t^{\alpha,a} &
- K_{2,2}^{b,\alpha} t^{\alpha,a} &
K_{2,1}^{b,\beta} t^{\beta,b} &
1-K_{2,2}^{b,\beta} t^{\beta,b}
\end{array} \right]
\left[\begin{array}{c}
G^{a,\beta}_{1,1} \\
G^{a,\beta}_{2,1} \\
G^{b,\beta}_{1,1} \\
G^{b,\beta}_{2,1} 
\end{array}\right]
=
\left[\begin{array}{c}
K^{a,\beta}_{1,1} \\
-K^{a,\beta}_{2,1} \\
K^{b,\beta}_{1,1} \\
-K^{b,\beta}_{2,1} 
\end{array}\right]
.
\end{equation}
The coefficients $K_{i,j}$ are of the form
$K_{i,j} = g_{i,i} t g_{i,j}$. For
instance,
$K_{i,j}^{a,\alpha} = g_{i,i}^{a,a}
t^{a,\alpha} g^{\alpha,\alpha}_{i,j}$.
The inversion of Eq.~(\ref{eq:Dy-4x4})
is described in section~\ref{sec:inv-F}
for two ferromagnetic electrodes having
a parallel spin orientation,
and in section~\ref{sec:inv-AF}
for two ferromagnetic electrodes
having an antiparallel spin orientation.

\subsection{Parallel magnetizations}
\label{sec:inv-F}
If $t^{a,\alpha} = t^{b,\beta}$, the
$4 \times 4$ Dyson matrix given
by Eq.~(\ref{eq:Dy-4x4}) can be written
in terms of $2 \times 2$ blocs:
\begin{equation}
\label{eq:DF-2x2}
\hat{\cal D}_{\rm F}
= \left[ \begin{array}{cc}
\HK_{\rm F} & \HL_{\rm F} \\
\HL^*_{\rm F} & \HK_{\rm F} \end{array} \right]
,
\end{equation}
where~$\HK_{\rm F}$ and~$\HL_{\rm F}$
can be obtained from Eq.~(\ref{eq:Dy-4x4}):
$$
\HK_{\rm F} =
\left[ \begin{array}{cc}
1-K_{1,1} & K_{1,2} \\
K_{2,1} & 1-K_{2,2}
\end{array}
\right] \mbox{ , and }
\HK_{\rm AF} =
\left[ \begin{array}{cc}
-L_{1,1} & L_{1,2} \\
L_{2,1} & -L_{2,2}
\end{array} \right]
.
$$
The inverse of $\hat{\cal D}_{\rm F}$ given
by~(\ref{eq:DF-2x2}) takes the form
$$
\hat{\cal D}_{\rm F}^{-1} =
\left[ \begin{array}{cc}
\HM_{\rm F}^{-1} & 
- \HK_{\rm F}^{-1} \HL_{\rm F} \HN_{\rm F}^{-1} \\
- \HK_{\rm F}^{-1} \HL_{\rm F}^* \HM_{\rm F}^{-1} &
\HN^{-1}_{\rm F} \end{array} \right]
,
$$
with $\HM_{\rm F}= \HK_{\rm F} -
\HL _{\rm F}\HK^{-1}_{\rm F} \HL^{*}_{\rm F}$
and $\HN_{\rm F} = \HK_{\rm F} -
\HL^*_{\rm F} \HK^{-1}_{\rm F} \HL_{\rm F}$.
The matrix in Eq.~(\ref{eq:DF-2x2}) can be evaluated
explicitly to obtain the
different terms in the Green's
function~(\ref{eq:G-beta-beta}):
\begin{eqnarray}
\label{eq:B4}
\LL \mbox{Im} \left[ g_{2,1}^{\beta,\alpha}
t^{\alpha,a} G_{1,1}^{a,\beta} \right] \RR
&=&- {1 \over 2}
\pi \rho_0^S f(\Delta_{\alpha,\beta})
\left[ - \frac{x_\uparrow}
{ 1 + x_\uparrow }
\left( \frac{ a_0}{R_{\alpha,\beta}} \right)^2
+
\frac{ x_\uparrow^2}
{ (1 + x_\uparrow )^2}
\frac{ a_0^3}{R_{\alpha,\beta} R_{\alpha',\beta}
R_{\alpha,\alpha'}} \right]\\
\label{eq:B5}
\LL \mbox{Im} \left[
- g_{2,2}^{\beta,\alpha} t^{\alpha,a}
G_{2,1}^{a,\beta} \right] \RR &=&
- \frac{t}{ (1-K_{1,1})^2 (1-K_{2,2})^2}
\tilde{g}_{2,2}^{\beta,\alpha} 
\left\{ -K_{2,1} (1-K_{1,1}) (1-K_{2,2})
\tilde{K}_{1,1}^{a,\beta} \right.\\
\nonb
 &-& \left.{1 \over 2} (1-K_{2,2})
(1-K_{1,1})^2 \tilde{K}_{2,1}^{a,\beta}
- \left[ K_{2,1} \left(
\tilde{L}_{1,1} (1-K_{2,2} )
+ \tilde{L}_{2,2} (1-K_{1,1}) \right) \right. \right. \\
\nonb
&+& \left. \left. {1 \over 2} \tilde{L}_{2,1}
(1-K_{1,1})(1-K_{2,2}) \right]
\tilde{K}_{1,1}^{b,\beta}
- {1 \over 2} \tilde{L}_{2,2}
(1-K_{1,1} )^2 \tilde{K}_{2,1}^{b,\beta}
\right\}\\
\label{eq:B6}
\LL \mbox{Im} \left[ g_{2,1}^{\beta,\alpha'}
t^{\alpha',a'} G^{a',\beta}_{1,1} \right] \RR
&=& - {1 \over 2} \pi \rho_0^S f(\Delta_{\alpha',\beta})
\left[ - \frac{x\uparrow }
{ 1 + x_\uparrow }
\left( \frac{ a_0}{R_{\alpha',\beta}} \right)^2
+
\frac{ x_\uparrow^2}
{ (1 + x_\uparrow^2}
\frac{ a_0^3}{R_{\alpha,\beta} R_{\alpha',\beta}
R_{\alpha,\alpha'}} \right]\\
\label{eq:B7}
\LL \mbox{Im} \left[ 
- g^{\beta,\alpha'}_{2,2} t^{\alpha',a'} G^{a',\beta}_{2,1}
\right] &=&
- \frac{t}{ (1-K_{1,1})^2 (1-K_{2,2})^2}
\tilde{g}^{\beta,\alpha'} \left\{ \left[
-K_{2,1} \left( \tilde{L}_{1,1} (1-K_{2,2})
+ \tilde{L}_{2,2} (1-K_{1,1}) \right) \right. \right.\\
\nonb
&-& \left. \left. {1 \over 2} \tilde{L}_{2,1} (1-K_{1,1})
(1-K_{2,2}) \right] 
\tilde{K}_{1,1}^{a,\beta}  
- {1 \over 2} \tilde{L}_{2,2} (1-K_{1,1})^2
\tilde{K}_{2,1}^{a,\beta} \right.\\
\nonb
&-& \left.K_{2,1} (1-K_{1,1})(1-K_{2,2})
\tilde{K}_{1,1}^{a',\beta}
- {1 \over 2} (1-K_{2,2}) (1-K_{1,1})^2
\tilde{K}_{2,1}^{a',\beta}
\right\}
,
\end{eqnarray}
where $x_\uparrow$ and~$x_\downarrow$
are given by Eqs.~(\ref{eq:x-up})
and~(\ref{eq:x-down}).
The function
$f(\Delta)$ is given by
$f(\Delta)=\Delta/\sqrt{\omega^2-\Delta^2}$.
The phase contribution has been factored out
in the coefficients $\tilde{L}_{i,j}$. For instance,
$L_{1,1} = \tilde{L}_{1,1} \exp{(i \varphi_{\alpha,\beta})}
\exp{(i \psi_{\alpha,\beta})}$, with
$\tilde{L}_{1,1} = - \pi^2 t^2 \rho^F_\uparrow
\rho^S_0 \left( a_0 / R_{\alpha,\beta}\right)$.
To obtain Eq.~(\ref{eq:P-Ferro})
we use a simplification of
Eqs.~(\ref{eq:B4})~--~(\ref{eq:B7}) in which
Eqs.~(\ref{eq:B4})~--~(\ref{eq:B7}) are transformed into
a local equation. This is done by replacing
$\Delta_{\alpha,\beta}$ and $\Delta_{\alpha',\beta}$
with $\Delta_{\beta}$. 

\subsection{Antiparallel magnetizations}
\label{sec:inv-AF}
The case of an antiparallel spin orientation of the
ferromagnetic electrodes can be treated in a similar
manner. Once the lines
and columns 3 and 4 have been interchanged,
the Dyson matrix given by Eq.~(\ref{eq:Dy-4x4})
takes the form
$$
\hat{\cal D}_{\rm AF} = \left[ \begin{array}{cc}
\HK_{\rm AF} & \HL_{\rm AF} \\
\HL_{\rm AF}^* & \HL_{\rm AF}^*
\end{array} \right]
,
$$
with
$$
\HK_{\rm AF} = \left[ \begin{array}{cc}
1-K_{1,1} & K_{1,2} \\
K_{2,1} & 1-K_{2,2} \end{array} \right]
\mbox{ , and }
\HL_{\rm AF} = \left[ \begin{array}{cc}
L_{1,2} & -L_{1,1} \\
-L_{2,2} & L_{2,1} \end{array} \right]
.
$$
The inverse is given by
$$
\hat{\cal D}_{\rm AF}^{-1}
= \left[ \begin{array}{cc}
\HM_{\rm AF}^{-1} & - \HK_{\rm AF}^{-1}
\HL_{\rm AF} (\HM_{\rm AF}^*)^{-1} \\
- (\HK^*_{\rm AF})^{-1} \HL_{\rm AF}^*
\HM_{\rm AF}^{-1} & 
(\HM_{\rm AF}^*)^{-1} \end{array}
\right]
,
$$
with $\HM_{\rm AF} = \HK_{\rm AF}-
\HL_{\rm AF} (\HK_{\rm AF}^*)^{-1}
\HL_{\rm AF}^*$.
The different terms in the Green's
function~(\ref{eq:G-beta-beta}) are the following:
\begin{eqnarray}
\label{eq:B8}
\LL \mbox{Im} \left[ g_{2,1}^{\beta,\alpha}
t^{\alpha,a} G_{1,1}^{a,\beta} \right] \RR
&=&-{1 \over 2} \pi \rho^S_0 f(\Delta_{\alpha,\beta})
\left[ - \frac{x_\uparrow}
{ 1 + x_\uparrow }
\left( \frac{ a_0}{R_{\alpha,\beta}} \right)^2
+
\frac{ x_\uparrow x_\downarrow}
{ (1 + x_\uparrow )(1+x_\downarrow)}
\frac{ a_0^3}{R_{\alpha,\beta} R_{\alpha',\beta}
R_{\alpha,\alpha'}} \right]
\\
\label{eq:B9}
\LL \mbox{Im} \left[
- g_{2,2}^{\beta,\alpha} t^{\alpha,a}
G_{2,1}^{a,\beta} \right] \RR &=&
-\frac{t}{ (1-K_{1,1}) (1-K_{2,2})}
\tilde{g}_{2,2}^{\beta,\alpha}
\left\{ -K_{2,1} \tilde{K}_{1,1}^{a,\beta}
-{1 \over 2} (1-K_{1,1}) \tilde{K}_{2,1}^{a,\beta}
- {1 \over 2} \tilde{L}_{2,2} \tilde{K}_{2,1}^{b,\beta}
\right.\\
\nonb
&-& \left.\frac{1}{1-K_{2,2}} \left[ 
K_{2,1} \tilde{L}_{1,1} + K_{1,2} \tilde{L}_{2,2}
+ {1 \over 2} \tilde{L}_{2,1} (1-K_{1,1}) \right]
\tilde{K}_{1,1}^{b,\beta}
\right\}
\\
\label{eq:B10}
\LL \mbox{Im} \left[ g_{2,1}^{\beta,\alpha'}
t^{\alpha',a'} G^{a',\beta}_{1,1} \right] \RR
&=&-{1 \over 2} \pi \rho^S_0 f(\Delta_{\alpha',\beta})
\left[ - \frac{x_\downarrow}
{ 1 + x_\downarrow }
\left( \frac{ a_0}{R_{\alpha',\beta}} \right)^2
+
\frac{ x_\uparrow x_\downarrow}
{ (1 + x_\uparrow )(1+x_\downarrow)}
\frac{ a_0^3}{R_{\alpha,\beta} R_{\alpha',\beta}
R_{\alpha,\alpha'}} \right]
\\
\label{eq:B11}
\LL \mbox{Im} \left[ 
- g^{\beta,\alpha'}_{2,2} t^{\alpha',a'} G^{a',\beta}_{2,1}
\right] \RR&=& \pi \rho^S_0
\frac{t}{ (1-K_{1,1})^2 (1-K_{2,2})^2} \left\{
(1-K_{2,2}) \left[ K_{1,2} \tilde{L}_{2,2} 
+ K_{2,1} \tilde{L}_{1,1}
+ {1 \over 2} \tilde{L}_{1,2} (1-K_{2,2}) \right]
\tilde{K}_{1,1}^{a,\beta} \right.\\
\nonb
&+& \left. {1 \over 2} \tilde{L}_{1,1}
(1-K_{1,1}) (1-K_{2,2}) \tilde{K}_{2,1}^{a,\beta}
+ {1 \over 2} (1-K_{1,1}) (1-K_{2,2})^2
\tilde{K}_{2,1}^{b,\beta} \right.\\
\nonb
&+& \left. K_{1,2} (1-K_{1,1}) (1-K_{2,2})
\tilde{K}_{1,1}^{b,\beta}
\right\}
.
\end{eqnarray}
Using a ``local'' approximation in which
$\Delta_{\alpha,\beta}$ and
$\Delta_{\alpha',\beta}$ are replaced by
$\Delta_\beta$ leads to Eq.~(\ref{eq:P-AntiFerro}).

\end{document}